\documentclass[sn-mathphys]{sn-jnl}

\usepackage{lineno}
\usepackage{overpic}
\jyear{2021}%

\theoremstyle{thmstyleone}%
\theoremstyle{thmstyletwo}%

\theoremstyle{thmstylethree}%

\begin{document}

\title[Effective Variable Depth Local Search for the BMCP]{Effective Variable Depth Local Search for the Budgeted Maximum Coverage Problem}


\author[]{\fnm{Jianrong} \sur{Zhou}}
\equalcont{These authors contributed equally to this work.}

\author[]{\fnm{Jiongzhi} \sur{Zheng}}
\equalcont{These authors contributed equally to this work. \vspace{-1em}}

\author[]{\fnm{Kun} \sur{He}$^*$}
\equalcont{The first two authors contribute equally to this work. \\$^*$: Corresponding author. Email: brooklet60@hust.edu.cn}

\affil[]{\orgdiv{School of Computer Science and Technology}, \orgname{Huazhong University of Science and Technology}, \orgaddress{\street{1037 Luoyu Road}, \city{Wuhan}, \postcode{430074}, \state{Hubei}, \country{China}\vspace{-1em}}}


\abstract{We address the Budgeted Maximum Coverage Problem (BMCP), which is a natural and more practical extension of the standard 0-1 knapsack problem and the set cover problem. Given $m$ elements with nonnegative weights, $n$ subsets of elements with nonnegative costs, and a total budget, BMCP aims to select some subsets such that the total cost of selected subsets does not exceed the budget, and the total weight of associated elements is maximized. In this paper, we propose a variable depth local search algorithm (VDLS) for the BMCP. VDLS first generates an initial solution by a greedy algorithm, then iteratively improves the solution through a partial depth-first search method, that can improve the solution by simultaneously changing the states (selected or not) of multiple subsets. Such method allows VDLS to explore the solution space widely and deeply, and to yield high-quality solutions. We further propose a neighbour structure to boost the algorithm performance, that is, both subsets have a neighbour relation if they share at least one common associated element. By applying the neighbour structure, VDLS can adjust the selected subsets while losing as few covered elements as possible. Since the existing BMCP benchmarks only have simple structures and small scales, we design 60 new instances with relatively large scales and complex structures to enrich the diversity of the BMCP instances. Experimental results on 30 public instances and 60 new instances we designed demonstrate that VDLS significantly outperforms the existing heuristic and the general CPLEX exact solver.
}



\vspace{-10em}
\maketitle

\section{Introduction}
The Budgeted Maximum Coverage Problem (BMCP)~\cite{Khuller1999} is a natural extension of the standard 0-1 knapsack problem~\cite{Kellerer2004} and the set cover problem~\cite{Balas1972}. Let $E=\{1,...,m\}$ be a set of $m$ elements where each element $j\in{E}$ has a nonnegative weight $w_j\geq{0}$ and $I=\{1,...,n\}$ be a set of $n$ items where each item $i\in{I}$ has a nonnegative cost $c_i\geq{0}$ and covers a subset of elements $E'_i\subseteq{E}$ associated by a relationship matrix $\mathbf{A}$. The goal of the BMCP is to select a collection $S\subseteq{I}$ that maximizes the sum of the weights of all the elements covered by the items in $S$, with the capacity constraint that the sum of the costs of the items in $S$ cannot exceed a given budget $L$.

The BMCP is closely related to two classical NP-hard problems, the 0-1 knapsack problem and the (weighted) set cover problem. If all the elements are unweighted, the BMCP corresponds to a variant of the weighted set cover problem~\cite{Nemhauser1978} that aims to select a collection $S\subseteq{I}$ and maximizes the total number of elements covered by items in $S$, with the capacity constraint that the total weight of the items in $S$ cannot exceed the budget $L$. Further, if all the items are also unweighted, the BMCP corresponds to a variant of the set cover problem~\cite{Hochba1997} of picking $L$ items such that the total number of covered elements is maximized. Also, the BMCP could degenerate into the 0-1 knapsack problem with weighted costs and weighted profits when each item covers exactly one element and each element is covered by exactly one item. 
Since both the NP-hard 0-1 knapsack problem and the NP-hard (weighted) set cover problem are reducible to the BMCP, the BMCP is NP-hard and is computationally challenging.


The BMCP has a wide range of real-world applications, such as service provider operations, location of network monitors~\cite{Suh2006}, news recommendation systems~\cite{Li2011}, worker employment, decision making~\cite{Jana2021,Jana2021CAM,Jana2022}, software package installation, facility location~\cite{Khuller1999}, and hybrid Software Defined Network optimization~\cite{Kar2016}. And there are some studies related to the extension problems of the BMCP, such as the generalized maximum coverage problem~\cite{Cohen2008}, the maximum coverage problem with group budget constraint~\cite{Chekuri2004}, the budgeted maximum coverage with overlapping costs~\cite{Curtis2010}, the ground-set-cost budgeted maximum coverage problem~\cite{Staereling2016}, etc. In summary, the BMCP is a significant and challenging NP-hard problem.

Moreover, the BMCP is highly related to the Set-Union Knapsack Problem (SUKP)~\cite{Goldschmidt1994}, wherein each item has a nonnegative profit and each element has a nonnegative weight. The goal of the SUKP is to select a subset $S\subseteq{I}$ that maximizes the total profits of the items in $S$, with the constraint that the sum of the weights of all the elements covered by the items in $S$ cannot exceed a given knapsack capacity $C$. In a word, the BMCP is a variant problem of the SUKP, and can be transferred to the SUKP by swapping the attributes of the items and elements~\cite{Li2021}. The SUKP has received much attention in recent years. Early in 1994, Goldschmidt et al. first presented an exact algorithm based on dynamic programming to solve the SUKP~\cite{Goldschmidt1994}. Later on, the approximation algorithms~\cite{Arulselvan2014,Taylor2016} and (meta) heuristic algorithms~\cite{He2018,Wei2019,Lin2019,Wu2020,Wei2021} were proposed for the SUKP.

However, as a highly related problem to the SUKP, there are relatively few researches on the standard BMCP after it was first presented by Khuller et al. in 1999~\cite{Khuller1999}. Khuller et al. presented a $(1-1/e)$-approximation algorithm, and claimed it is the best possible approximation unless NP $\subseteq$ DTIME$(n^{\text{O(loglog} n)})$. Later on, some approximation algorithms~\cite{Cohen2008,Staereling2016,Piva2019} and two greedy constructive algorithms~\cite{Kar2016} were proposed for various extensions or variants of the BMCP. Recently in 2021, Li et al.~\cite{Li2021} proposed a meta-heuristic algorithm based on tabu search and reinforcement learning, called probability learning based tabu search (PLTS), for the standard BMCP problem. 

To the best of our knowledge, PLTS is the only practical algorithm proposed for solving the standard BMCP. Like the effective heuristic algorithms for the SUKP \cite{Wei2019,Wei2021}, PLTS follows the tabu search framework and explores the solution space by very simple search operators. Such a framework with simple search operators might result in a relatively small search scope and thus lead the algorithm not well exploring the solution space and easy to get stuck in local optima. 

To handle this issue and make up for the lack of efficient heuristics for solving the BMCP, which is a challenging NP-hard problem with widely practical applications, in this work we aim to present an effective and novel heuristic algorithm for the BMCP. Note that our method is not based on the widely adopted tabu search framework, yet could significantly improve the performance of the existing heuristic algorithms and enrich the solutions. Specifically, we propose a variable depth local search algorithm, denoted as VDLS. The VDLS first generates an initial solution by the greedy constructive algorithm proposed by Khuller et al.~\cite{Khuller1999}, then improves the solution by a partial depth-first tree search method that the number of the branch nodes and the depth of the search tree are restricted. As the initial solution guarantees the approximation ratio of $(1-1/e)$, the improved solution also guarantees the approximation ratio of $(1-1/e)$. 

The partial depth-first search method used in the VDLS enables to explore the solution space with a deep and wide search scope, that can improve the current solution by simultaneously changing the states (selected or not) of multiple items.
During the search, we define an effective neighbour structure to improve the performance of the depth-first search method by cutting relatively low-quality branch nodes. Specifically, a neighbour item of item $i$ must cover at least one element covered by item $i$. With the help of the neighbour structure, VDLS can adjust the selected items while losing as few covered elements as possible, so as to explore the solution space more efficiently. Since the search depth and search scope of the partial depth-first search method used in the VDLS are much larger than those of the PLTS, and the neighbour structure used in the VDLS is effective, our VDLS algorithm significantly outperforms the PLTS for solving the BMCP.

Li et al.~\cite{Li2021} also proposed the first set of BMCP benchmarks to facilitate the evaluation of algorithms.
However, the structure of this set of BMCP benchmarks is simple and their problem scale is small in 585 to 1000 for the number of items/elements. To this end, we design 60 new instances in larger scales (1000 to 5200 for the number of items/elements) and have some latent complex structure. 
Then, we compare the proposed VDLS algorithm with the best-performing BMCP heuristic of the PLTS algorithm, as well as the general CPLEX solver, with 2 hours and 5 hours of time limit, on the existing 30 benchmarks and the 60 new instances we designed. The results show that our proposed VDLS algorithm significantly outperforms the PLTS algorithm and the CPLEX solver for solving the BMCP.


The main contributions of this work are as follows:

\begin{itemize}
\item We propose an effective algorithm called the VDLS based on the partial depth-first tree search method for solving the NP-hard BMCP. VDLS can explore the solution space widely and deeply, so as to yield high-quality solutions. Our method suggests a new solving perspective other than tabu search to solve the BMCP.
\item We define an effective neighbour structure for the BMCP, which can improve the performance of the proposed local search algorithm. The neighbour structure we define is general and may be applied to improve other local search algorithms for solving this problem.
\item As the existing BMCP benchmarks only have simple structures and in small scales, we design 60 new BMCP instances in larger scales and with complex structural features to enrich the diversity of the BMCP instances. 
\item Extensive experiments demonstrate that VDLS significantly outperforms the existing heuristic PLTS, as well as the CPLEX exact solver, for solving the BMCP. The improvement of VDLS over PLTS is more significant on our designed 60 instances, indicating that VDLS is more effective for solving large and complex BMCP instances, and our designed instances can distinguish different algorithms' performance better.

\end{itemize}

The rest of this paper is organized as follows. Section~\ref{sec_Pre} introduces basic conceptions and formal definitions about the BMCP. Section~\ref{sec_Alg} describes our proposed approach, including the main framework of the VDLS, the greedy constructive algorithm, and the local search process based on a partial depth-first search tree, and further presents the advantages of the proposed VDLS algorithm. Section~\ref{sec_Exp} presents experimental results of the proposed VDLS algorithm and the comparison with the existing heuristic. Section~\ref{sec_Con} contains the concluding remarks.

\section{Problem Formulation}
\label{sec_Pre}
Given a set of $m$ elements $E = \{1, ..., m\}$ where each element $j \in E$ has a nonnegative weight $w_{j} \geq 0$, a set of $n$ items $I = \{1, ..., n\}$ where each item $i \in I$ has a nonnegative cost $c_{i} \geq 0$, a binary relationship matrix $\mathbf{A} \in \{0, 1\}^{\mathrm{m} \times \mathrm{n}}$ where $\mathbf{A}_{ji} = 1$ indicates that element $j$ is covered by item $i$ (otherwise $\mathbf{A}_{ji} = 0$), and a budget $L$. Let $S$ be a subset of $I$, we define $C(S)$ be the total cost of all the items in $S$, and $W(S)$ be the total weight of the elements covered by all the items in $S$, respectively. The goal of the BMCP is to find a subset of the items, $S \subseteq I$, that maximizes $W(S)$ and satisfies the budget constraint, i.e., $C(S) \leq L$.

Let $\mathbf{x} \in \{0, 1\}^{\mathrm{n}}$ be a binary vector that represents a solution $S$ of the BMCP where $\mathbf{x}_{i} = 1$ if item $i \in S$ (otherwise $\mathbf{x}_{i} = 0$), $\mathbf{c} \in \mathbb{N}^{\mathrm{n}}$ be a vector of $n$ items where $c_{i}$ is the cost of item $i$.
The BMCP can be formulated as the following integer linear programming problem: 

\vspace{-1em}
\begin{equation*}
 \mathrm{Maximize} \quad W(S) = \sum\limits_{j = 1}^m w_j [y_j > 0]
\end{equation*}
\begin{equation*}
\textit{s.t.}~~~~(1) \quad \mathbf{y} = \mathbf{A} \mathbf{x}
\end{equation*}
\begin{equation*}
~~~~~~~~~~~~~~~~~~~~~(2) \quad C(S) = \mathbf{c}^{\mathrm{T}} \mathbf{x} \leqslant L,
\end{equation*}
where $[\cdot]$ is the Iverson bracket that $[P]=1$ when statement $P$ is true, otherwise $[P]=0$. 



\begin{algorithm}[b]
\caption{VDLS($k, d, T$)}
\label{alg_VDLS}
\setlength\parindent{1em} \textbf{Input:} the maximum search width $k$, the maximum search depth $d$, the
\par\setlength\parindent{4.5em} cut-off time $T$ 
\par\setlength\parindent{1em} \textbf{Output:} a solution $S$ 
\par\setlength\parindent{1.5em} 1: $S := \mathrm{Greedy}()$ 
\par\setlength\parindent{1.5em} 2: Initialize the current search depth $d' := 0$ 
\par\setlength\parindent{1.5em} 3: Initialize the set of the visited items $R := \emptyset$ 
\par\setlength\parindent{1.5em} 4: Initialize $step := 0$, $i := 0$ 
\par\setlength\parindent{1.5em} 5: $P :=$ a random permutation of $\{1,...,n\}$ 
\par\setlength\parindent{1.5em} 6:  \textbf{while} the cut-off time $T$ is not reached \textbf{do} 
\par\setlength\parindent{1.5em} 7: \qquad \textbf{if} flipping $P_i$ in $S$ is feasible \textbf{then} 
\par\setlength\parindent{1.5em} 8: \qquad \qquad Initialize the input solution $S_{in} := S$
\par\setlength\parindent{1.5em} 9: \qquad \qquad $S' := \mathrm{Local\_Search}(S_{in}, k, d, d', S, P_i, R)$
\par\setlength\parindent{1em} 10: \qquad \qquad \textbf{if} $W(S') > W(S)$ \textbf{then} 
\par\setlength\parindent{1em} 11: \qquad \qquad \qquad \setlength\parindent{3em} $S := S'$, $step := 0$
\par\setlength\parindent{1em} 12: \qquad \qquad \textbf{end if}
\par\setlength\parindent{1em} 13: \qquad \textbf{end if} 
\par\setlength\parindent{1em} 14: \qquad $i := i \% n + 1$, $step := step + 1$
\par\setlength\parindent{1em} 15: \qquad \textbf{if} $step \geq n$ \textbf{then} 
\par\setlength\parindent{1em} 16: \qquad \qquad \textbf{break}
\par\setlength\parindent{1em} 17: \qquad \textbf{end if} 
\par\setlength\parindent{1em} 18: \textbf{end while} 
\par\setlength\parindent{1em} 19:  \textbf{return} $S$
\end{algorithm}

\section{The Proposed VDLS Algorithm}
\label{sec_Alg}
The proposed variable depth local search (VDLS) algorithm consists of an initialization stage and an improvement stage. In the initialization stage, we generate an initial solution by an approximation algorithm~\cite{Khuller1999}, which is actually a greedy constructive method. In the improvement stage, we propose a partial depth-first tree search method to improve the initial solution. The branch size of each node and the depth of the entire depth-first search tree are restricted. The search depth is variable and controlled by an early-stop strategy, which can improve the efficiency of the algorithm.

This section first introduces the main process of the VDLS, then describes the greedy constructive algorithm and the local search process in the VDLS, respectively. Finally we present the advantages of the proposed algorithm.

\subsection{The Main Process}
The main procedure of the VDLS is shown in Algorithm~\ref{alg_VDLS}. The VDLS algorithm first generates an initial solution $S$ by a greedy constructive algorithm (line 1), then improves the solution by a local search process based on the partial depth-first search approach until the stopping criterion is met (lines 6-18). At each iteration, the partial depth-first search approach (i.e., the LocalSearch function) tries to improve the current solution from a selected item ($P_i$ in line 7) as the root of the search tree, until a better solution is found or the entire search tree has been traversed under the restriction. 
The method of selecting item as the root of the search tree is to randomly traverse all the items in $I$. Specifically, VDLS first generates a random permutation of $\{1, 2, ..., n\}$, denoted as $P$ (line 5). Then, the algorithm sequentially traverses each item in $P$ as the root of the search tree (line 14).


The VDLS algorithm stops when the cut-off time $T$ is reached (line 6) or the partial depth-first search process starting from any item $i \in I$ (i.e., the root of the search tree) cannot improve the current solution (lines 15-17).

\begin{algorithm}[b]
\caption{Greedy()}
\label{alg_greedy}
\setlength\parindent{1em} \textbf{Output:} a solution $S_{out}$
\par\setlength\parindent{1.5em} 1: $S_{out} := \emptyset$
\par\setlength\parindent{1.5em} 2: \textbf{while} TRUE \textbf{do}
\par\setlength\parindent{1.5em} 3: \qquad $U := \{i \mid i \in I, i \not\in S_{out}, C(S_{out}) + c_i \leq L\}$
\par\setlength\parindent{1.5em} 4: \qquad \textbf{if} $U = \emptyset$ \textbf{then}
\par\setlength\parindent{1.5em} 5: \qquad \qquad \textbf{break}
\par\setlength\parindent{1.5em} 6: \qquad \textbf{end if}
\par\setlength\parindent{1.5em} 7: \qquad $p := \mathop{\arg\max}_{p \in U} \frac{W(S_{out} \cup \{p\}) - W(S_{out})}{c_p}$
\par\setlength\parindent{1.5em} 8: \qquad $S_{out} := S_{out} \cup \{p\}$
\par\setlength\parindent{1.5em} 9: \textbf{end while}
\par\setlength\parindent{1em} 10: $q := \mathop{\arg\max}_{i \in I, c_i \leq L} W(\{i\})$
\par\setlength\parindent{1em} 11: \textbf{if} $W(S_{out}) < W(\{q\})$ \textbf{then}
\par\setlength\parindent{1em} 12: \qquad $S_{out} := \{q\}$
\par\setlength\parindent{1em} 13: \textbf{end if}
\par\setlength\parindent{1em} 14:  \textbf{return} $S_{out}$
\end{algorithm}

\subsection{The Greedy Constructive Algorithm} 
\label{subsec_greedy}

We then introduce the initialization method in VDLS. Khuller et al.~\cite{Khuller1999} proposed a simple greedy algorithm and further proposed an approximation algorithm based on this greedy algorithm. We adopt this greedy algorithm, 
as depicted in Algorithm~\ref{alg_greedy}, to construct the initial solution. 

This greedy algorithm outputs the best among two candidate solutions (lines 11-14).
The first candidate is generated by a greedy heuristic (lines 1-9) that iteratively adds an item $i \in I \setminus S_{out}$ into $S_{out}$ that satisfies the total cost of all items in $S_{out} \cup \{i\}$ not exceeding the budget $L$ and the increased weight per unit cost of item $i$ is maximized.
The second candidate is a single item $q$ with the maximized weight $W(\{q\})$ (line 10). 

The second option is necessary, as in some special cases, the second candidate is better than the first one. For example, $I = \{1, 2\}$ contains two items with $c_1 = 1, c_2 = L$ (the budget $L>2$), $E = \{1, 2\}$ contains two elements with $w_1 = 1, w_2 = L-1$, element 1 is covered by item 1 and element 2 is covered by item 2. In this case, the solution contains either item 1 or item 2. The first candidate will choose item 1 since the ratio of the total weight of the elements covered by item 1 to the cost of item 1 
is greater than that of item 2 (i.e., $w_1/c_1 > w_2/c_2$). And the second candidate will choose item 2 since the total weight of the covered elements of item 2 is greater than that of item 1. Obviously, in the above example, the second candidate yields a better solution.

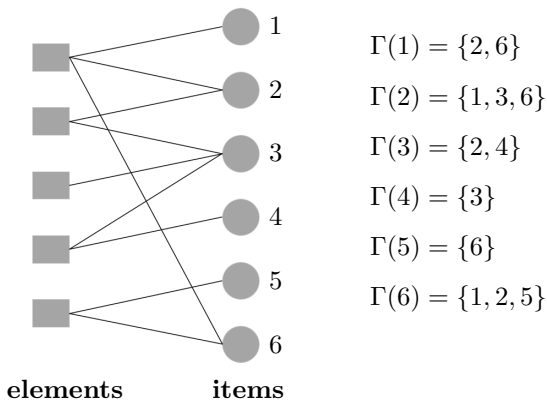
\begin{figure}[b]
\centering
\begin{overpic}[width=0.7\columnwidth]{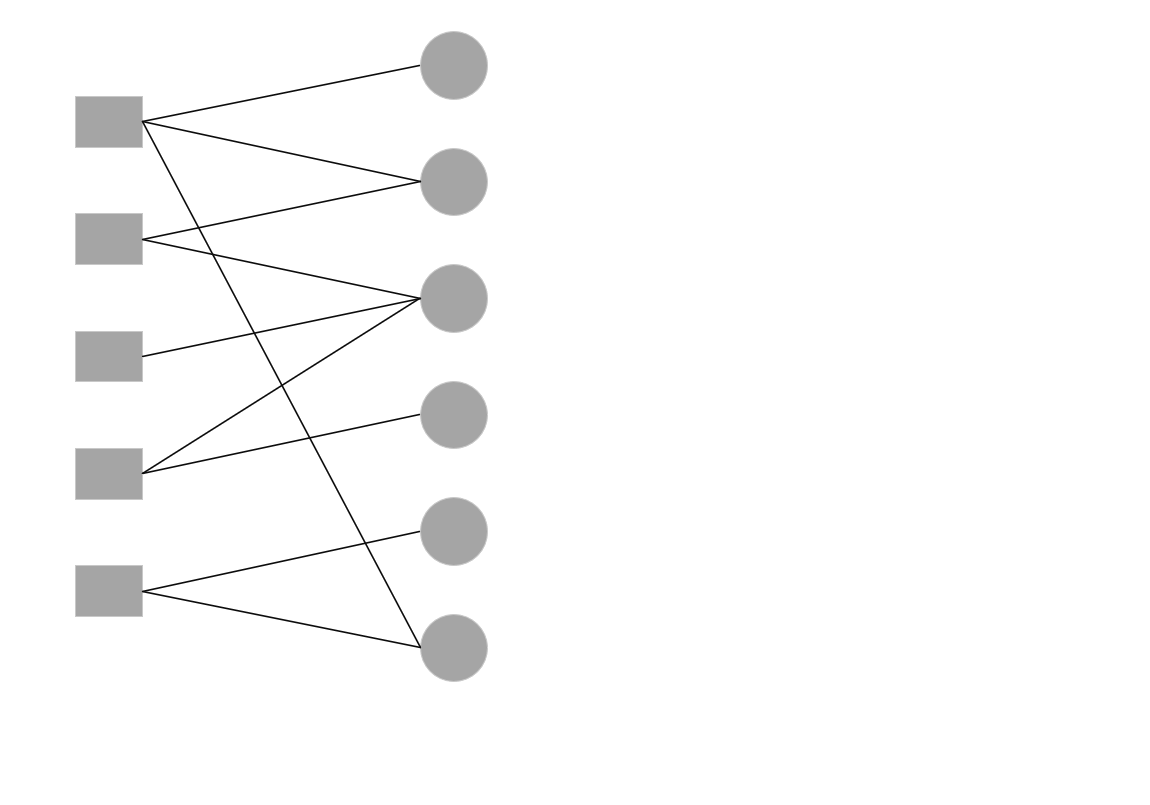}
\put(2.5, 5){\textbf{elements}}
\put(35, 5){\textbf{items}}
\put(44, 63.3){$1$}
\put(44, 53.1){$2$}
\put(44, 42.9){$3$}
\put(44, 32.7){$4$}
\put(44, 22.5){$5$}
\put(44, 12.3){$6$}
\put(60, 60){$\Gamma(1) = \{2, 6\}$}
\put(60, 52){$\Gamma(2) = \{1, 3, 6\}$}
\put(60, 44){$\Gamma(3) = \{2, 4\}$}
\put(60, 36){$\Gamma(4) = \{3\}$}
\put(60, 28){$\Gamma(5) = \{6\}$}
\put(60, 20){$\Gamma(6) = \{1, 2, 5\}$}
\end{overpic}
\caption{
The neighbour structure of a BMCP instance with 5 elements and 6 items. The edges represent the relationship between the elements and the items. The neighbours of each item are presented on the right.
}
\vspace{-1em}
\label{fig_neighbour}
\end{figure}

\begin{algorithm}[b]
\caption{Local\_Search($S_{in}, k, d, d', S', p, R$)}
\label{alg_LS}
\setlength\parindent{1em} \textbf{Input:} the input solution $S_{in}$, the maximum search width $k$, the maximum 
\par\setlength\parindent{4.5em} search depth $d$, current search depth $d'$, current solution $S$, flipping 
\par\setlength\parindent{4.5em} item $p$, set of the visited items $R$
\par\setlength\parindent{1em} \textbf{Output:} solution $S_{out}$
\par\setlength\parindent{1.5em} 1: $S' := flip(S, p)$
\par\setlength\parindent{1.5em} 2: \textbf{if} $W(S') > W(S_{in})$ \textbf{then}
\par\setlength\parindent{1.5em} 3: \qquad \textbf{return} $S'$
\par\setlength\parindent{1.5em} 4: \textbf{end if}
\par\setlength\parindent{1.5em} 5: \textbf{if} $d' \geq d$ \textbf{then}
\par\setlength\parindent{1.5em} 6: \qquad \textbf{return} $\emptyset$
\par\setlength\parindent{1.5em} 7: \textbf{end if}
\par\setlength\parindent{1.5em} 8: $U := \{i \mid i \in \Gamma(p), i \notin R, C(flip(S', i)) \leq L\}$ 
\par\setlength\parindent{1.5em} 9: Sort $U$ in descending order of the $gain(S', i)$, $i \in U$
\par\setlength\parindent{1em} 10: \textbf{for} $j := 1 : \min(|U|, k)$ \textbf{do}
\par\setlength\parindent{1em} 11: \qquad $q_j := $ $j$-th item in $U$
\par\setlength\parindent{1em} 12: \qquad $R := R \cup q_j$
\par\setlength\parindent{1em} 13: \qquad $S_{temp} := \mathrm{Local\_Search}(S_{in}, k, d, d'+1, S', q_j, R)$
\par\setlength\parindent{1em} 14: \qquad \textbf{if} $S_{temp} \neq \emptyset$ \textbf{then}
\par\setlength\parindent{1em} 15: \qquad \qquad \textbf{return} $S_{temp}$
\par\setlength\parindent{1em} 16: \qquad \textbf{end if}
\par\setlength\parindent{1em} 17: \textbf{end for}
\par\setlength\parindent{1em} 18: \textbf{return} $\emptyset$
\end{algorithm}

\subsection{Variable Depth Local Search}

Before introducing the local search procedure, we first present several definitions. 

We define $Cover(i) = \{j \mid \mathbf{A}_{ji} = 1\}$ as a set of elements covered by item $i$. 
We then define $\Gamma(i) = \{i' \mid i' \in I, i' \neq i, Cover(i) \cap Cover(i') \neq \emptyset\}$ as a set of items that cover at least one element in $Cover(i)$ (i.e., any item $i' \in \Gamma(i)$ have at least one common covered element with item $i$). We also call the items in $\Gamma(i)$ the neighbours of item $i$. Figure~\ref{fig_neighbour} shows a BMCP instances with 5 elements and 6 items, and presents the neighbours of each item.


We define a function $flip(S, i)$ 
that changes the state of item $i$ in solution $S$, i.e., the flipping operator removes item $i$ from $S$ if $i \in S$, otherwise it adds item $i$ into $S$. And function $flip(S, i)$ returns a solution $S'$ obtained by flipping item $i$ in $S$. The flipping is the basic operation in the entire VDLS algorithm. 
Moreover, we define the function $gain(S, i)$ as the difference of the objective value after flipping item $i$ in $S$. The function $gain(S, i)$ is defined as follows:
\begin{equation} 
    \label{eq_gain}
    gain(S, i) = 
    \begin{cases}
        W(S \cup \{i\}) - W(S), & i \notin S \\
        W(S - \{i\}) - W(S), & i \in S \\
    \end{cases}
\end{equation}


The local search process in VDLS actually applies a partial depth-first search to improve the input solution $S_{in}$. The key is how to decide the branch nodes. Algorithm~\ref{alg_LS} shows the local search procedure, which tries to find a solution better than the input solution $S_{in}$ by flipping a selected item $p$ (i.e., the branch node) in the current solution $S$ recursively (line 13). The branch node $p$ is decided according to a search tree that the maximum depth and the maximum width are restricted to $d$ and $k$, respectively.
The root of the search tree is a randomly selected item $P_i \in I$ (see lines 5, 7, 9 in Algorithm~\ref{alg_VDLS}).

After selecting the root of the search tree, the algorithm starts the partial depth-first search process. At each searching node, the algorithm first flips item $p$ in $S$ to obtain the current solution $S'$. To decide the branch nodes at the next level after flipping $p$, the algorithm constructs a candidate set $U$ consisting of the items in $\Gamma(p)$ that have not been visited in the current search tree (i.e., not belonging to $R$, a set that records the visited items in the current search tree) and flipping any of them in the current solution $S'$ is feasible (line 8). Then the algorithm sorts the candidate set $U$ in descending order of $gain(S', i), i \in U$ (line 9), since flipping an item with a larger value of $gain$ leads to a higher-quality solution. After sorting $U$, the algorithm selects the top $k$ items from $U$ as the branch nodes at the next level (line 10).

Moreover, during each partial depth-first search process, an early-stop strategy is applied to improve the efficiency. That is, if the current solution $S'$ is better than the input solution $S_{in}$, then the algorithm terminates the process and returns the current solution $S'$ (lines 2-4). 

We further analyze the time complexity of each iteration of VDLS (see line 9 in Algorithm~\ref{alg_VDLS}). As depicted in Algorithm~\ref{alg_LS}, the local search process traverses a search tree where the maximum width and the maximum depth are limited by two hyperparameters, $k$ and $d$. Thus the algorithm actually traverses $O(k^d)$ nodes in this process. In the local search process, the branching strategy is based on the neighbor structure. Therefore, the time complexity of expanding each search node is impacted by the neighbor structure of VDLS, and we give a complexity analysis on the random instances as follows.

Let $\alpha = \frac{1}{nm}\sum_{i=1}^{n}\sum_{j=1}^{m}\mathbf{A}_{ji}$ represents the density of the relationship matrix.
The expectation of the number of elements covered by an item is $\alpha m$, so the time complexity of the $flip$ operator (see lines 1, 8 in Algorithm~\ref{alg_LS}) is $O(\alpha m)$.
The probability of two items covering at least one common element is $\sigma = 1 - (1 - \alpha^2)^m$, and the expectation size of $\Gamma(i)$ is $\sigma (n - 1)$.
Obtaining the list $U$ (see lines 8, 9 in Algorithm~\ref{alg_LS}) is the main time cost of expanding each search node. The process of obtaining list $U$ consists of flipping each item in $\Gamma(i)$ and sorting the order according to $gain$ (Eq. \ref{eq_gain}). Therefore, the time complexity of obtaining the list $U$ is $O(\alpha \sigma mn + \sigma n log(\sigma n)))$, which is also the time complexity of expanding each search node.
Finally, the time complexity of each iteration of VDLS is $O(k^d(\alpha \sigma mn + \sigma n log(\sigma n)))$ on the random instances.

\subsection{Advantages of the VDLS Algorithm}
This subsection introduces the advantages of the proposed VDLS algorithm, including the effective neighbour structure we defined for the BMCP, and the search neighbourhood of the partial depth-first search method.

\subsubsection{Effective Neighbour Structure}
We first explain why the proposed neighbour structure is reasonable and effective for the BMCP. Note that once an item $i$ is flipped, it will affect the quality of the items that have common covered elements, i.e., the neighbours of item $i$ ($\Gamma(i)$). Specifically, if item $i$ is removed from $S$, then the elements covered by item $i$ but not covered by $S\setminus\{i\}$ become uncovered in the current solution. Thus the next potentially high-quality move might be to select another item to cover these uncovered elements, which must be a neighbour of item $i$, and vice versa. Therefore, it is reasonable and effective to select the candidate items from the neighbours $\Gamma(i)$ of item $i$ for the next step after flipping item $i$ at the current step. By applying the neighbours as the candidate set, the search scope is significantly reduced, and the efficiency of the local search process is improved. 

\subsubsection{High-quality Search Space}
The VDLS algorithm has a better neighbourhood search space than the existing BMCP heuristic algorithm, the PLTS algorithm~\cite{Li2021}, which only has two neighbourhood operators, $flip$ and $swap$. The $flip$ operator in the PLTS is the same as the function $flip(S, i)$ in our proposed VDLS algorithm, that changes the state of an item in the current solution. And the $swap$ operator removes an item from the current solution $S$ and adds another item into $S$. These two operators are simple and are actually belonging to the special cases of our neighbourhood operator, i.e., the search space of the PLTS is equal to that of the VDLS with the maximum depth $d = 2$ and the maximum width $k = n$ of the search tree. 

In the proposed VDLS algorithm, we do a deeper search with a larger $d$ ($d = 8$ by default) and shrink the search width of the branch nodes in the next level of the search tree, by only considering the neighbour items that have common covered elements with the item flipped in the current search node. To further improve the search efficiency, we only consider the top $k$ ($k = 7$ by default) neighbours with a larger value of gain (Eq. \ref{eq_gain}) as the candidates of the branch nodes. Moreover, the early-stop strategy can improve the efficiency of the algorithm by terminating in advance. 

In summary, the VDLS algorithm has a higher-quality search space than the PLTS algorithm, since the VDLS can improve the solution by flipping at most $d$ items continuously during each search step, and the candidates of the branch nodes are refined by the neighbour structure and the restriction of the search width. 

\section{Experimental Results}
\label{sec_Exp}
In this section, we present the experimental results, including the comparison of VDLS with different settings of parameters ($d$ and $k$), the comparison of VDLS with the existing BMCP heuristic and the general CPLEX solver, and the ablation study on the neighbour structure and branching strategy of the VDLS algorithm. We test the algorithms on the existing 30 BMCP benchmarks and 60 new instances in relatively large scales and with complex structural features that we designed. We first present the benchmark instances, the experimental setup, and then present the experimental results and discussions.

\subsection{Benchmark Instances}
For the BMCP, there are a set of 30 public BMCP instances~\footnote{https://github.com/lly53/BMCP\_instance.} proposed by Li et al. in 2021~\cite{Li2021}. Among these 30 instances, the number of items or elements (i.e., $n$ or $m$) ranges from 585 to 1,000, the budget of each instance $L$ is either 1500 or 2000, and the density of relationship matrix $\alpha = \frac{1}{mn}\sum\nolimits_{i = 1}^{n}{\sum\nolimits_{j = 1}^{m}{\mathbf{A}_{ji}}}$ is 0.075 (resp. 0.05) for the instances with $L = 1500$ (resp. $L = 2000$). 

The existing BMCP benchmarks are in rather small scales ($n, m$ in [500, 1000]) and built with simple structures. The relationship matrix of each instance is determined by randomly selecting $\alpha \times n \times m$ edges with replacement between $n$ items and $m$ elements, and then setting $\mathbf{A}_{ji}$ equals to 1 (resp. 0) if edge $(j,i)$ is selected (resp. not selected). 

To enrich the diversity of the BMCP instances, we propose two sets with a total of 60 new BMCP instances in larger scale ($n, m$ in [1000, 1600], or [4000, 5200]) and with complex structural features. Given the number of the items and elements, i.e., $n$ and $m$, the relationship matrix of each proposed instance is determined by repeating the following process $t$ ($t = 3$ by default) times. First, we evenly divide both items and elements into $g$ ($g = 25$ by default) groups. Then for the $l$-th ($l \in \{1,2,...,g\}$) pair of groups consisting of the $l$-th item group with $n_l$ items and $l$-th element group with $m_l$ elements, we randomly select $\rho \times n_l \times m_l$ edges with replacement to determine the relationship matrix. $\rho = \frac{1}{m_l n_l}\sum\nolimits_{i = 1}^{n_l}{\sum\nolimits_{j = 1}^{m_l}{\mathbf{A}_{n_l^i m_l^j}}}$ is the density of the relationship matrix between the items and elements belonging to the $l$-th pair of groups, where $n_l^i$ and $m_l^j$ are the $i$-th item in the $l$-th item group and the $j$-th element in the $l$-th element group respectively. 

The proposed 60 instances can be divided into two groups according to the scale. The first group contains 30 instances with the number of items or elements ranging from 1000 to 1600, the budget $L$ either 2000 or 3000, and the density parameter $\rho$ is 0.5 (resp. 0.3) for the instances with $L =$ 2000 (resp. 3000). Note that the density parameter $\rho$ in our proposed instances is larger than the density parameter $\alpha$ in the existing instances proposed by Li et al.~\cite{Li2021}, because the density in each pair of item and element groups will be diluted when considering all the items and elements. The second group contains 30 instances with the number of the items or elements ranging from 4000 to 5200, the budget $L$ either 7000 or 10000, and the density parameter $\rho$ is 0.5 (resp. 0.3) for the instances with $L =$ 7000 (resp. 10000). The proposed 60 new instances are available at https://github.com/JHL-HUST/VDLS/.

In summary, we test our algorithm on three sets of BMCP instances. The first set consists of the 30 instances proposed by Li et al.~\cite{Li2021}. The second (resp. third) set consists of the 30 new instances with the number of items or elements ranging from 1000 to 1600 (resp. 4000 to 5200). The three sets of instances are listed at the first column of Table~\ref{table1-10} (or Table~\ref{table1-30}), Table~\ref{table2} and Table~\ref{table3}, respectively, where an instance named $bmcp\_n\_m\_\alpha\_L$ in Table~\ref{table1-10} or Table~\ref{table1-30} indicates an instance of the first set with $n$ items, $m$ elements, the density of the total relationship matrix $\alpha$, and the budget $L$, and an instance named $bmcp\_n\_m\_\rho\_L$ in Table~\ref{table2} or Table~\ref{table3} indicates an instance of the second set or the third set with $n$ items, $m$ elements, the density of the relationship matrix in each pair of item and element groups $\rho$, and the budget $L$.



\subsection{Experimental Setup}
We compare our VDLS algorithm with the general CPLEX solver in solving the first set of instances (since it is too time consuming for the CPLEX solver to obtain high-quality lower and upper bounds when solving the instances of the second and the third sets), and compare the VDLS with the best-performing BMCP heuristic, the PLTS algorithm~\cite{Li2021} in solving all the three sets with a total of 90 instances. Each instance is calculated 10 times by each algorithm.

For the CPLEX solver (version 12.8), we employ it under the time limit of 2 hours as well as 5 hours for each instance based on the 0/1 integer linear programming model. The CPLEX solver finds the upper bound and lower bound of the 30 instances in the first set. For the heuristics, both the VDLS and the PLTS run 10 times for each instance with different random seeds independently. Since Li et al.~\cite{Li2021} test the PLTS under the time limit of 10 minutes, we first compare the VDLS and the PLTS with 10 minutes of time limit for each run. And we further compare these two algorithms under the time limit of 30 minutes to evaluate their performance with a longer cut-off time and do a more comprehensive comparison.

The experiments of the VDLS were coded in the C++ programming language and complied by g++ with the ’-O3’ option. The experiments of all the algorithms, including CPLEX, PLTS, VDLS and its variants, were performed on a Linux server with Intel® Xeon® E5-2650 v3 2.30 GHz 10-core CPU and 256G RAM. The parameters of the VDLS include the maximum depth $d$ and maximum width $k$ of the search tree. The comparison of the VDLS with different settings of these two parameters is presented in the following subsection.


\begin{table}[t]
\centering
\caption{Comparison of VDLS with different settings of parameters with 10 or 30 minutes of time limit in solving the 30 instances of the first set. Regions of the parameters are $d \in [5,10]$ and $k \in [5,10]$. Results are expressed by the average value of the best solutions of all the 30 instances obtained by the algorithms in 10 runs. Best results appear in bold.}
\label{table_para}
\begin{tabular}{lrrrrrr} \toprule
       & $k=5$            & $k=6$            & $k=7$            & $k=8$            & $k=9$            & $k=10$  \\ \hline
\multicolumn{7}{l}{10 minutes of time limit}                                                                    \\ \hline
$d=5$  & 92191.2          & 92178.4          & 92174.3          & 92180.6          & 92178.1          & 92211.0 \\
$d=6$  & 92180.8          & 92219.7          & 92258.8          & 92249.2          & 92241.6          & 92264.6 \\
$d=7$  & 92243.1          & 92257.5          & 92254.7          & \textbf{92290.5} & 92272.6          & 92264.4 \\
$d=8$  & \textbf{92290.5} & \textbf{92290.5} & \textbf{92290.5} & 92270.9          & 92273.7          & 92222.0 \\
$d=9$  & 92280.9          & 92252.1          & 92173.4          & 92106.3          & 91985.0          & 91905.2 \\
$d=10$ & 92286.1          & 92173.3          & 92009.0          & 91925.7          & 91843.6          & 91744.3 \\ \hline
\multicolumn{7}{l}{30 minutes of time limit}                                                                    \\ \hline
$d=5$  & 92191.2          & 92178.4          & 92174.3          & 92180.6          & 92178.1          & 92211.0 \\
$d=6$  & 92180.8          & 92219.7          & 92258.8          & 92249.2          & 92241.6          & 92264.6 \\
$d=7$  & 92243.1          & 92257.5          & 92254.7          & \textbf{92290.5} & \textbf{92290.5} & 92277.3 \\
$d=8$  & \textbf{92290.5} & \textbf{92290.5} & \textbf{92290.5} & \textbf{92290.5} & \textbf{92290.5} & 92290.1 \\
$d=9$  & \textbf{92290.5} & 92268.1          & 92268.4          & 92243.1          & 92184.4          & 92109.7 \\
$d=10$ & \textbf{92290.5} & 92271.4          & 92206.0          & 92109.0          & 92016.8          & 91929.8 \\ \bottomrule
\end{tabular}
\end{table}

\begin{table}[b]
\centering
\caption{Comparison of the PLTS and VDLS with both 10 minutes of time limit, and the CPLEX solver with 2 hours and 5 hours of time limit in solving the 30 instances of the first set. Best results appear in bold.}
\label{table1-10}
\scalebox{0.56}{\begin{tabular}{lrrrrrrrrrrr} \toprule
\multirow{2}{*}{Instances}    & \multicolumn{2}{r}{CPLEX(2 hours)} & \multicolumn{1}{l}{} & \multicolumn{2}{r}{CPLEX(5 hours)} & \multicolumn{1}{l}{} & \multicolumn{2}{r}{PLTS}  &  & \multicolumn{2}{r}{VDLS}    \\ \cline{2-3} \cline{5-6} \cline{8-9} \cline{11-12} 
                              & LB              & UB               & \multicolumn{1}{l}{} & LB              & UB               & \multicolumn{1}{l}{} & best           & average  &  & best             & average  \\ \hline
bmcp\_585\_600\_0.05\_2000    & 67910           & 73495.9          &                      & 70742           & 74224.9          &                      & 70997          & 70892.6  &  & \textbf{71102}   & 71034.4  \\
bmcp\_585\_600\_0.075\_1500   & 68418           & 77549.4          &                      & 69172           & 76716.9          &                      & 70677          & 70535.9  &  & \textbf{71025}   & 70492.5  \\
bmcp\_600\_585\_0.05\_2000    & 66184           & 71739.7          &                      & 66452           & 71094.3          &                      & 67187          & 66807.1  &  & \textbf{67636}   & 67621.6  \\
bmcp\_600\_585\_0.075\_1500   & 68145           & 77321.5          &                      & 70113           & 76332.9          &                      & 70318          & 70318.0  &  & \textbf{70588}   & 70277.0  \\
bmcp\_600\_600\_0.05\_2000    & 67917           & 73495.1          &                      & 68477           & 72880.9          &                      & 68453          & 68346.6  &  & \textbf{68738}   & 68580.7  \\
bmcp\_600\_600\_0.075\_1500   & 70947           & 77380.0          &                      & 71018           & 76337.7          &                      & 71746          & 71746.0  &  & \textbf{71904}   & 71780.7  \\
bmcp\_685\_700\_0.05\_2000    & 79997           & 88954.3          &                      & 80783           & 88447.7          &                      & 80111          & 79979.9  &  & \textbf{81227}   & 80482.5  \\
bmcp\_685\_700\_0.075\_1500   & 80443           & 92328.3          &                      & 81639           & 91379.0          &                      & 82955          & 82933.4  &  & \textbf{83286}   & 82886.6  \\
bmcp\_700\_685\_0.05\_2000    & 76139           & 83561.9          &                      & 77176           & 82815.1          &                      & \textbf{78054} & 77460.8  &  & \textbf{78054}   & 77742.9  \\
bmcp\_700\_685\_0.075\_1500   & 75841           & 86956.9          &                      & 76033           & 86566.8          &                      & \textbf{78869} & 78580.1  &  & \textbf{78869}   & 78642.9  \\
bmcp\_700\_700\_0.05\_2000    & 76367           & 85587.3          &                      & 77056           & 84855.4          &                      & 77424          & 77225.8  &  & \textbf{78458}   & 78180.4  \\
bmcp\_700\_700\_0.075\_1500   & 81645           & 93026.7          &                      & 81645           & 92152.0          &                      & 84570          & 83406.5  &  & \textbf{84576}   & 84576.0  \\
bmcp\_785\_800\_0.05\_2000    & 90705           & 102198.7         &                      & 91319           & 101585.7         &                      & 92505          & 92080.7  &  & \textbf{92740}   & 92414.2  \\
bmcp\_785\_800\_0.075\_1500   & 92358           & 107354.5         &                      & 92358           & 106842.7         &                      & 94245          & 94243.8  &  & \textbf{95221}   & 95221.0  \\
bmcp\_800\_785\_0.05\_2000    & 86139           & 97891.1          &                      & 86813           & 97477.3          &                      & 88562          & 88424.3  &  & \textbf{89138}   & 89054.7  \\
bmcp\_800\_785\_0.075\_1500   & 89018           & 103644.1         &                      & 89229           & 102868.0         &                      & 91021          & 90888.9  &  & \textbf{91856}   & 91722.9  \\
bmcp\_800\_800\_0.05\_2000    & 90344           & 101141.0         &                      & 89872           & 100373.8         &                      & 90951          & 90697.4  &  & \textbf{91795}   & 91572.8  \\
bmcp\_800\_800\_0.075\_1500   & 94049           & 108713.0         &                      & 94049           & 108005.6         &                      & 95533          & 95508.8  &  & \textbf{95995}   & 95568.5  \\
bmcp\_885\_900\_0.05\_2000    & 100085          & 114313.8         &                      & 99845           & 113747.0         &                      & 101293         & 100857.0 &  & \textbf{102277}  & 102059.2 \\
bmcp\_885\_900\_0.075\_1500   & 102423          & 122684.4         &                      & 102933          & 122093.5         &                      & 105718         & 105707.7 &  & \textbf{106940}  & 106446.9 \\
bmcp\_900\_885\_0.05\_2000    & 97088           & 110828.8         &                      & 96945           & 109948.5         &                      & 98840          & 98485.5  &  & \textbf{99590}   & 99365.1  \\
bmcp\_900\_885\_0.075\_1500   & 99881           & 119045.5         &                      & 99888           & 118554.8         &                      & 103596         & 103223.1 &  & \textbf{105141}  & 105024.8 \\
bmcp\_900\_900\_0.05\_2000    & 100108          & 115682.6         &                      & 100412          & 114551.1         &                      & 101265         & 101210.7 &  & \textbf{102055}  & 101402.6 \\
bmcp\_900\_900\_0.075\_1500   & 101035          & 120452.8         &                      & 99888           & 118554.8         &                      & 104521         & 104521.0 &  & \textbf{105081}  & 104919.1 \\
bmcp\_985\_1000\_0.05\_2000   & 107820          & 125903.7         &                      & 107488          & 124487.4         &                      & 109523         & 108681.8 &  & \textbf{110669}  & 109792.6 \\
bmcp\_985\_1000\_0.075\_1500  & 110769          & 134440.4         &                      & 111177          & 133789.8         &                      & 113541         & 113439.4 &  & \textbf{115505}  & 114765.3 \\
bmcp\_1000\_985\_0.05\_2000   & 108714          & 126779.1         &                      & 110134          & 125574.8         &                      & 111212         & 110379.1 &  & \textbf{112057}  & 111583.1 \\
bmcp\_1000\_985\_0.075\_1500  & 108801          & 131873.0         &                      & 108801          & 130981.4         &                      & 112250         & 112160.2 &  & \textbf{113615}  & 113096.1 \\
bmcp\_1000\_1000\_0.05\_2000  & 109928          & 131194.2         &                      & 111155          & 128583.6         &                      & 111829         & 111318.6 &  & \textbf{113331}  & 113124.8 \\
bmcp\_1000\_1000\_0.075\_1500 & 115313          & 139152.9         &                      & 115824          & 137900.4         &                      & 118244         & 118243.1 &  & \textbf{120246}  & 119994.5 \\ \hline
Average                       & 89484.4         & 103156.3         &                      & 89947.9         & 102324.1         &                      & 91533.7        & 91276.8  &  & \textbf{92290.5} & 91980.9 
      \\ \bottomrule              
\end{tabular}}
\end{table}

\begin{table}[t]
\centering
\caption{Comparison of the PLTS and VDLS with both 30 minutes of time limit, and the CPLEX solver with 2 hours and 5 hours of time limit in solving the 30 instances of the first set. Best results appear in bold.}
\label{table1-30}
\scalebox{0.56}{\begin{tabular}{lrrrrrrrrrrr} \toprule
\multirow{2}{*}{Instances}    & \multicolumn{2}{r}{CPLEX(2 hours)} & \multicolumn{1}{l}{} & \multicolumn{2}{r}{CPLEX(5 hours)} & \multicolumn{1}{l}{} & \multicolumn{2}{r}{PLTS}   &  & \multicolumn{2}{r}{VDLS}                       \\ \cline{2-3} \cline{5-6} \cline{8-9} \cline{11-12} 
                              & LB              & UB               & \multicolumn{1}{l}{} & LB              & UB               & \multicolumn{1}{l}{} & best            & average  &  & best             & \multicolumn{1}{r}{average} \\ \hline
bmcp\_585\_600\_0.05\_2000    & 67910           & 73495.9          &                      & 70742           & 74224.9          &                      & \textbf{71102}  & 70915.0  &  & \textbf{71102}   & 71034.4                     \\
bmcp\_585\_600\_0.075\_1500   & 68418           & 77549.4          &                      & 69172           & 76716.9          &                      & 70677           & 70670.2  &  & \textbf{71025}   & 70492.5                     \\
bmcp\_600\_585\_0.05\_2000    & 66184           & 71739.7          &                      & 66452           & 71094.3          &                      & \textbf{67636}  & 67041.5  &  & \textbf{67636}   & 67621.6                     \\
bmcp\_600\_585\_0.075\_1500   & 68145           & 77321.5          &                      & 70113           & 76332.9          &                      & 70318           & 70318.0  &  & \textbf{70588}   & 70277.0                     \\
bmcp\_600\_600\_0.05\_2000    & 67917           & 73495.1          &                      & 68477           & 72880.9          &                      & 68453           & 68386.8  &  & \textbf{68738}   & 68580.7                     \\
bmcp\_600\_600\_0.075\_1500   & 70947           & 77380.0          &                      & 71018           & 76337.7          &                      & 71746           & 71746.0  &  & \textbf{71904}   & 71780.7                     \\
bmcp\_685\_700\_0.05\_2000    & 79997           & 88954.3          &                      & 80783           & 88447.7          &                      & 80197           & 80096.6  &  & \textbf{81227}   & 80482.5                     \\
bmcp\_685\_700\_0.075\_1500   & 80443           & 92328.3          &                      & 81639           & 91379.0          &                      & 82955           & 82944.2  &  & \textbf{83286}   & 82886.6                     \\
bmcp\_700\_685\_0.05\_2000    & 76139           & 83561.9          &                      & 77176           & 82815.1          &                      & 77876           & 77372.7  &  & \textbf{78054}   & 77742.9                     \\
bmcp\_700\_685\_0.075\_1500   & 75841           & 86956.9          &                      & 76033           & 86566.8          &                      & \textbf{78869}  & 78810.4  &  & \textbf{78869}   & 78642.9                     \\
bmcp\_700\_700\_0.05\_2000    & 76367           & 85587.3          &                      & 77056           & 84855.4          &                      & 78028           & 77501.5  &  & \textbf{78458}   & 78200.7                     \\
bmcp\_700\_700\_0.075\_1500   & 81645           & 93026.7          &                      & 81645           & 92152.0          &                      & 84570           & 83720.7  &  & \textbf{84576}   & 84576.0                     \\
bmcp\_785\_800\_0.05\_2000    & 90705           & 102198.7         &                      & 91319           & 101585.7         &                      & 92431           & 92195.7  &  & \textbf{92740}   & 92419.7                     \\
bmcp\_785\_800\_0.075\_1500   & 92358           & 107354.5         &                      & 92358           & 106842.7         &                      & 94245           & 94245.0  &  & \textbf{95221}   & 95221.0                     \\
bmcp\_800\_785\_0.05\_2000    & 86139           & 97891.1          &                      & 86813           & 97477.3          &                      & 88562           & 88560.3  &  & \textbf{89138}   & 89054.7                     \\
bmcp\_800\_785\_0.075\_1500   & 89018           & 103644.1         &                      & 89229           & 102868.0         &                      & 91021           & 90971.8  &  & \textbf{91856}   & 91722.9                     \\
bmcp\_800\_800\_0.05\_2000    & 90344           & 101141.0         &                      & 89872           & 100373.8         &                      & 91748           & 90934.2  &  & \textbf{91795}   & 91604.5                     \\
bmcp\_800\_800\_0.075\_1500   & 94049           & 108713.0         &                      & 94049           & 108005.6         &                      & 95533           & 95533.0  &  & \textbf{95995}   & 95568.5                     \\
bmcp\_885\_900\_0.05\_2000    & 100085          & 114313.8         &                      & 99845           & 113747.0         &                      & 101293          & 101142.0 &  & \textbf{102277}  & 102059.2                    \\
bmcp\_885\_900\_0.075\_1500   & 102423          & 122684.4         &                      & 102933          & 122093.5         &                      & 106364          & 105782.6 &  & \textbf{106940}  & 106446.9                    \\
bmcp\_900\_885\_0.05\_2000    & 97088           & 110828.8         &                      & 96945           & 109948.5         &                      & 98840           & 98645.4  &  & \textbf{99590}   & 99368.7                     \\
bmcp\_900\_885\_0.075\_1500   & 99881           & 119045.5         &                      & 99888           & 118554.8         &                      & \textbf{105141} & 103846.9 &  & \textbf{105141}  & 105024.8                    \\
bmcp\_900\_900\_0.05\_2000    & 100108          & 115682.6         &                      & 100412          & 114551.1         &                      & 101265          & 101211.4 &  & \textbf{102055}  & 101481.4                    \\
bmcp\_900\_900\_0.075\_1500   & 101035          & 120452.8         &                      & 99888           & 118554.8         &                      & 104521          & 104521.0 &  & \textbf{105081}  & 104919.1                    \\
bmcp\_985\_1000\_0.05\_2000   & 107820          & 125903.7         &                      & 107488          & 124487.4         &                      & 109567          & 108938.0 &  & \textbf{110669}  & 109792.6                    \\
bmcp\_985\_1000\_0.075\_1500  & 110769          & 134440.4         &                      & 111177          & 133789.8         &                      & 113541          & 113541.0 &  & \textbf{115505}  & 114765.3                    \\
bmcp\_1000\_985\_0.05\_2000   & 109928          & 131194.2         &                      & 111155          & 128583.6         &                      & 111176          & 110712.6 &  & \textbf{112057}  & 111624.7                    \\
bmcp\_1000\_985\_0.075\_1500  & 115313          & 139152.9         &                      & 115824          & 137900.4         &                      & 112250          & 112218.9 &  & \textbf{113615}  & 113096.1                    \\
bmcp\_1000\_1000\_0.05\_2000  & 108714          & 126779.1         &                      & 110134          & 125574.8         &                      & 112640          & 111577.8 &  & \textbf{113331}  & 113129.6                    \\
bmcp\_1000\_1000\_0.075\_1500 & 108801          & 131873.0         &                      & 108801          & 130981.4         &                      & 119453          & 118499.3 &  & \textbf{120246}  & 119994.5                    \\ \hline
Average                       & 89484.4         & 103156.3         &                      & 89947.9         & 102324.1         &                      & 91733.9         & 91420.0  &  & \textbf{92290.5} & 91987.1                    \\ \bottomrule
\end{tabular}}
\end{table}

\begin{table}[t]
\centering
\caption{Comparison results of VDLS and PLTS, with 10 or 30 minutes of time limit, on the 30 instances of the second set. Best results appear in bold.}
\label{table2}
\scalebox{0.65}{\begin{tabular}{lrrrrr} \toprule
\multirow{2}{*}{Instances}  & \multicolumn{2}{r}{PLTS}   &  & \multicolumn{2}{r}{VDLS}     \\ \cline{2-3} \cline{5-6} 
                            & best            & average  &  & best              & average  \\ \hline
\multicolumn{6}{l}{10 minutes of time limit}                                               \\ \hline
bmcp\_1100\_1000\_0.3\_3000 & 142574          & 141985.1 &  & \textbf{143475}   & 143222.1 \\
bmcp\_1100\_1000\_0.5\_2000 & 125790          & 125479.5 &  & \textbf{126582}   & 125970.4 \\
bmcp\_1100\_1100\_0.3\_3000 & 154726          & 153867.0 &  & \textbf{157530}   & 156462.5 \\
bmcp\_1100\_1100\_0.5\_2000 & 138160          & 137490.0 &  & \textbf{139141}   & 138694.5 \\
bmcp\_1100\_1200\_0.3\_3000 & 169980          & 168661.8 &  & \textbf{172660}   & 172075.3 \\
bmcp\_1100\_1200\_0.5\_2000 & 149402          & 148618.4 &  & \textbf{150493}   & 149860.2 \\
bmcp\_1200\_1100\_0.3\_3000 & \textbf{161848} & 159267.2 &  & \textbf{161848}   & 161817.3 \\
bmcp\_1200\_1100\_0.5\_2000 & 136073          & 135347.1 &  & \textbf{137700}   & 136564.8 \\
bmcp\_1200\_1200\_0.3\_3000 & 169219          & 168710.1 &  & \textbf{172595}   & 172095.7 \\
bmcp\_1200\_1200\_0.5\_2000 & 148706          & 147717.1 &  & \textbf{151440}   & 149480.5 \\
bmcp\_1200\_1300\_0.3\_3000 & 184156          & 181891.2 &  & \textbf{184384}   & 183583.0 \\
bmcp\_1200\_1300\_0.5\_2000 & 160487          & 159181.1 &  & \textbf{162154}   & 160955.4 \\
bmcp\_1300\_1200\_0.3\_3000 & 168304          & 167883.9 &  & \textbf{170786}   & 169724.7 \\
bmcp\_1300\_1200\_0.5\_2000 & 150249          & 148937.0 &  & \textbf{151402}   & 150995.1 \\
bmcp\_1300\_1300\_0.3\_3000 & 180593          & 179909.5 &  & \textbf{184699}   & 183180.9 \\
bmcp\_1300\_1300\_0.5\_2000 & 160677          & 159351.7 &  & \textbf{162146}   & 161493.6 \\
bmcp\_1300\_1400\_0.3\_3000 & 195341          & 195308.4 &  & \textbf{199134}   & 198818.9 \\
bmcp\_1300\_1400\_0.5\_2000 & 172244          & 171175.5 &  & \textbf{173954}   & 173718.2 \\
bmcp\_1400\_1300\_0.3\_3000 & 181154          & 179794.4 &  & \textbf{183987}   & 183792.4 \\
bmcp\_1400\_1300\_0.5\_2000 & 161658          & 159182.6 &  & \textbf{162544}   & 162190.5 \\
bmcp\_1400\_1400\_0.3\_3000 & 199067          & 197227.7 &  & \textbf{200136}   & 199593.2 \\
bmcp\_1400\_1400\_0.5\_2000 & 173693          & 172899.6 &  & \textbf{175324}   & 173898.9 \\
bmcp\_1400\_1500\_0.3\_3000 & 210450          & 208051.0 &  & \textbf{212492}   & 211196.8 \\
bmcp\_1400\_1500\_0.5\_2000 & 183146          & 182427.8 &  & \textbf{186818}   & 186118.5 \\
bmcp\_1500\_1400\_0.3\_3000 & 193676          & 192820.3 &  & \textbf{197219}   & 196642.5 \\
bmcp\_1500\_1400\_0.5\_2000 & 171949          & 170581.2 &  & \textbf{174284}   & 173461.7 \\
bmcp\_1500\_1500\_0.3\_3000 & 208630          & 206404.9 &  & \textbf{211108}   & 210222.4 \\
bmcp\_1500\_1500\_0.5\_2000 & 183599          & 181456.4 &  & \textbf{187052}   & 184938.1 \\
bmcp\_1500\_1600\_0.3\_3000 & \textbf{226408} & 221501.2 &  & \textbf{226408}   & 224514.1 \\
bmcp\_1500\_1600\_0.5\_2000 & 197566          & 195828.2 &  & \textbf{198663}   & 198137.0 \\ \hline
Average                     & 171984.2        & 170631.9 &  & \textbf{173938.6} & 173114.0 \\ \hline
\multicolumn{6}{l}{30 minutes of time limit}                                               \\ \hline
bmcp\_1100\_1000\_0.3\_3000 & 142835          & 142428.7 &  & \textbf{143475}   & 143224.6 \\
bmcp\_1100\_1000\_0.5\_2000 & 125974          & 125790.0 &  & \textbf{126917}   & 126021.6 \\
bmcp\_1100\_1100\_0.3\_3000 & 154936          & 154661.7 &  & \textbf{157530}   & 156462.5 \\
bmcp\_1100\_1100\_0.5\_2000 & \textbf{139196} & 138169.0 &  & 139141            & 138719.4 \\
bmcp\_1100\_1200\_0.3\_3000 & 169113          & 168763.6 &  & \textbf{172660}   & 172075.3 \\
bmcp\_1100\_1200\_0.5\_2000 & 150301          & 149366.4 &  & \textbf{150493}   & 149944.0 \\
bmcp\_1200\_1100\_0.3\_3000 & \textbf{161848} & 159620.3 &  & \textbf{161848}   & 161848.0 \\
bmcp\_1200\_1100\_0.5\_2000 & 135858          & 135509.2 &  & \textbf{137700}   & 136831.2 \\
bmcp\_1200\_1200\_0.3\_3000 & 169219          & 168821.7 &  & \textbf{172595}   & 172151.8 \\
bmcp\_1200\_1200\_0.5\_2000 & 149654          & 148679.1 &  & \textbf{151440}   & 149610.0 \\
bmcp\_1200\_1300\_0.3\_3000 & 183004          & 182220.4 &  & \textbf{186083}   & 183599.0 \\
bmcp\_1200\_1300\_0.5\_2000 & 160369          & 159735.4 &  & \textbf{162154}   & 161146.9 \\
bmcp\_1300\_1200\_0.3\_3000 & 168586          & 168473.8 &  & \textbf{170786}   & 169724.7 \\
bmcp\_1300\_1200\_0.5\_2000 & 151067          & 150034.7 &  & \textbf{151402}   & 151049.4 \\
bmcp\_1300\_1300\_0.3\_3000 & 180608          & 179978.4 &  & \textbf{184699}   & 183239.0 \\
bmcp\_1300\_1300\_0.5\_2000 & 160485          & 159511.3 &  & \textbf{162146}   & 161536.7 \\
bmcp\_1300\_1400\_0.3\_3000 & 197709          & 195681.9 &  & \textbf{199134}   & 198889.4 \\
bmcp\_1300\_1400\_0.5\_2000 & 173349          & 171665.0 &  & \textbf{173954}   & 173799.2 \\
bmcp\_1400\_1300\_0.3\_3000 & 182881          & 180143.5 &  & \textbf{183987}   & 183796.3 \\
bmcp\_1400\_1300\_0.5\_2000 & 161995          & 160647.0 &  & \textbf{162544}   & 162312.9 \\
bmcp\_1400\_1400\_0.3\_3000 & 199879          & 197229.0 &  & \textbf{200136}   & 199616.7 \\
bmcp\_1400\_1400\_0.5\_2000 & 174683          & 173393.1 &  & \textbf{175583}   & 173984.9 \\
bmcp\_1400\_1500\_0.3\_3000 & 210077          & 208315.6 &  & \textbf{212492}   & 211264.4 \\
bmcp\_1400\_1500\_0.5\_2000 & 184259          & 183289.9 &  & \textbf{186818}   & 186250.3 \\
bmcp\_1500\_1400\_0.3\_3000 & 194156          & 193554.8 &  & \textbf{197219}   & 196703.4 \\
bmcp\_1500\_1400\_0.5\_2000 & 173475          & 171426.9 &  & \textbf{174284}   & 173576.4 \\
bmcp\_1500\_1500\_0.3\_3000 & 209929          & 207795.3 &  & \textbf{211108}   & 210393.7 \\
bmcp\_1500\_1500\_0.5\_2000 & 184082          & 182676.2 &  & \textbf{187052}   & 185221.6 \\
bmcp\_1500\_1600\_0.3\_3000 & \textbf{226408} & 221641.3 &  & \textbf{226408}   & 224926.6 \\
bmcp\_1500\_1600\_0.5\_2000 & 198214          & 196659.0 &  & \textbf{199678}   & 198462.6 \\ \hline
Average                     & 172471.6        & 171196.1 &  & \textbf{174048.9} & 173212.8 \\ \bottomrule
\end{tabular}}
\end{table}

\begin{table}[t]
\centering
\caption{Comparison results of VDLS and PLTS, with 10 or 30 minutes of time limit, on the 30 instances of the third set. Best results appear in bold.}
\label{table3}
\scalebox{0.65}{\begin{tabular}{lrrrrr} \toprule
\multirow{2}{*}{Instances}   & \multicolumn{2}{r}{PLTS} &  & \multicolumn{2}{r}{VDLS}       \\ \cline{2-3} \cline{5-6} 
                             & best        & average    &  & best               & average   \\ \hline
\multicolumn{6}{l}{10 minutes of time limit}                                                \\ \hline
bmcp\_4200\_4000\_0.3\_10000 & 1775432     & 1736438.9  &  & \textbf{1835947}   & 1827846.2 \\
bmcp\_4200\_4000\_0.5\_7000  & 1692056     & 1662063.5  &  & \textbf{1735128}   & 1720514.4 \\
bmcp\_4200\_4200\_0.3\_10000 & 1856948     & 1824810.6  &  & \textbf{1924499}   & 1912643.9 \\
bmcp\_4200\_4200\_0.5\_7000  & 1759417     & 1735178.9  &  & \textbf{1801634}   & 1796598.8 \\
bmcp\_4200\_4400\_0.3\_10000 & 1964294     & 1928237.3  &  & \textbf{2021037}   & 2015973.6 \\
bmcp\_4200\_4400\_0.5\_7000  & 1846157     & 1815658.3  &  & \textbf{1893350}   & 1883659.3 \\
bmcp\_4400\_4200\_0.3\_10000 & 1817604     & 1797114.8  &  & \textbf{1914670}   & 1905583.0 \\
bmcp\_4400\_4200\_0.5\_7000  & 1762870     & 1724377.3  &  & \textbf{1807499}   & 1800460.9 \\
bmcp\_4400\_4400\_0.3\_10000 & 1948267     & 1918148.4  &  & \textbf{2026134}   & 2017604.4 \\
bmcp\_4400\_4400\_0.5\_7000  & 1817416     & 1806039.0  &  & \textbf{1893155}   & 1880770.0 \\
bmcp\_4400\_4600\_0.3\_10000 & 2016242     & 1989001.0  &  & \textbf{2096767}   & 2086931.0 \\
bmcp\_4400\_4600\_0.5\_7000  & 1913938     & 1895993.0  &  & \textbf{1985446}   & 1973491.3 \\
bmcp\_4600\_4400\_0.3\_10000 & 1939966     & 1894543.2  &  & \textbf{2005469}   & 1995349.8 \\
bmcp\_4600\_4400\_0.5\_7000  & 1822083     & 1803727.3  &  & \textbf{1891961}   & 1886326.4 \\
bmcp\_4600\_4600\_0.3\_10000 & 2027745     & 1996365.4  &  & \textbf{2103062}   & 2093885.6 \\
bmcp\_4600\_4600\_0.5\_7000  & 1932145     & 1886756.8  &  & \textbf{1977544}   & 1967468.6 \\
bmcp\_4600\_4800\_0.3\_10000 & 2090814     & 2062236.5  &  & \textbf{2199773}   & 2188595.1 \\
bmcp\_4600\_4800\_0.5\_7000  & 2006997     & 1964777.6  &  & \textbf{2059987}   & 2051531.7 \\
bmcp\_4800\_4600\_0.3\_10000 & 2008566     & 1974395.8  &  & \textbf{2088967}   & 2080467.4 \\
bmcp\_4800\_4600\_0.5\_7000  & 1902428     & 1879771.0  &  & \textbf{1975756}   & 1966373.7 \\
bmcp\_4800\_4800\_0.3\_10000 & 2071733     & 2034925.5  &  & \textbf{2180765}   & 2168243.6 \\
bmcp\_4800\_4800\_0.5\_7000  & 1981902     & 1955184.4  &  & \textbf{2063921}   & 2053363.9 \\
bmcp\_4800\_5000\_0.3\_10000 & 2221536     & 2157486.8  &  & \textbf{2274851}   & 2269242.4 \\
bmcp\_4800\_5000\_0.5\_7000  & 2055213     & 2032723.9  &  & \textbf{2139652}   & 2128574.9 \\
bmcp\_5000\_4800\_0.3\_10000 & 2076374     & 2039438.7  &  & \textbf{2179552}   & 2166424.9 \\
bmcp\_5000\_4800\_0.5\_7000  & 1987069     & 1962373.3  &  & \textbf{2057220}   & 2049177.5 \\
bmcp\_5000\_5000\_0.3\_10000 & 2154362     & 2103199.9  &  & \textbf{2255867}   & 2245749.3 \\
bmcp\_5000\_5000\_0.5\_7000  & 2077170     & 2041834.9  &  & \textbf{2146446}   & 2134731.1 \\
bmcp\_5000\_5200\_0.3\_10000 & 2235890     & 2211542.6  &  & \textbf{2347464}   & 2337836.8 \\
bmcp\_5000\_5200\_0.5\_7000  & 2119204     & 2099244.2  &  & \textbf{2218955}   & 2205056.8 \\ \hline
Average                      & 1962727.9   & 1931119.6  &  & \textbf{2036749.3} & 2027015.9 \\ \hline
\multicolumn{6}{l}{30 minutes of time limit}                                                \\ \hline
bmcp\_4200\_4000\_0.3\_10000 & 1778013     & 1747411.9  &  & \textbf{1835947}   & 1833108.2 \\
bmcp\_4200\_4000\_0.5\_7000  & 1685141     & 1666673.3  &  & \textbf{1738553}   & 1729188.0 \\
bmcp\_4200\_4200\_0.3\_10000 & 1874190     & 1839780.1  &  & \textbf{1925706}   & 1918624.2 \\
bmcp\_4200\_4200\_0.5\_7000  & 1764277     & 1742121.7  &  & \textbf{1811649}   & 1801880.2 \\
bmcp\_4200\_4400\_0.3\_10000 & 1954069     & 1932251.0  &  & \textbf{2023093}   & 2020209.8 \\
bmcp\_4200\_4400\_0.5\_7000  & 1853677     & 1820444.4  &  & \textbf{1903515}   & 1889809.5 \\
bmcp\_4400\_4200\_0.3\_10000 & 1833465     & 1816879.1  &  & \textbf{1920418}   & 1911396.2 \\
bmcp\_4400\_4200\_0.5\_7000  & 1746064     & 1730783.9  &  & \textbf{1810031}   & 1805200.7 \\
bmcp\_4400\_4400\_0.3\_10000 & 1945218     & 1922134.2  &  & \textbf{2027801}   & 2024036.1 \\
bmcp\_4400\_4400\_0.5\_7000  & 1844031     & 1821467.8  &  & \textbf{1901941}   & 1888300.5 \\
bmcp\_4400\_4600\_0.3\_10000 & 2003959     & 1993946.2  &  & \textbf{2100925}   & 2093610.1 \\
bmcp\_4400\_4600\_0.5\_7000  & 1937066     & 1906956.1  &  & \textbf{1989582}   & 1979416.6 \\
bmcp\_4600\_4400\_0.3\_10000 & 1930498     & 1913501.0  &  & \textbf{2008258}   & 2001368.8 \\
bmcp\_4600\_4400\_0.5\_7000  & 1845046     & 1817697.2  &  & \textbf{1898602}   & 1889658.5 \\
bmcp\_4600\_4600\_0.3\_10000 & 2027132     & 2005096.7  &  & \textbf{2106938}   & 2100124.5 \\
bmcp\_4600\_4600\_0.5\_7000  & 1921261     & 1900376.3  &  & \textbf{1980462}   & 1973677.2 \\
bmcp\_4600\_4800\_0.3\_10000 & 2120120     & 2091722.5  &  & \textbf{2201977}   & 2197572.4 \\
bmcp\_4600\_4800\_0.5\_7000  & 1989454     & 1967943.3  &  & \textbf{2070913}   & 2057914.3 \\
bmcp\_4800\_4600\_0.3\_10000 & 2019033     & 1991673.5  &  & \textbf{2096747}   & 2087482.0 \\
bmcp\_4800\_4600\_0.5\_7000  & 1922466     & 1901157.6  &  & \textbf{1991409}   & 1973721.9 \\
bmcp\_4800\_4800\_0.3\_10000 & 2100773     & 2058921.6  &  & \textbf{2183525}   & 2175464.8 \\
bmcp\_4800\_4800\_0.5\_7000  & 2009129     & 1973993.6  &  & \textbf{2066487}   & 2059458.7 \\
bmcp\_4800\_5000\_0.3\_10000 & 2184445     & 2157793.0  &  & \textbf{2280317}   & 2273869.0 \\
bmcp\_4800\_5000\_0.5\_7000  & 2087596     & 2050741.0  &  & \textbf{2146413}   & 2136313.2 \\
bmcp\_5000\_4800\_0.3\_10000 & 2069979     & 2051607.2  &  & \textbf{2183483}   & 2175171.2 \\
bmcp\_5000\_4800\_0.5\_7000  & 2013921     & 1977310.7  &  & \textbf{2066000}   & 2053359.9 \\
bmcp\_5000\_5000\_0.3\_10000 & 2144223     & 2126462.9  &  & \textbf{2261620}   & 2251527.0 \\
bmcp\_5000\_5000\_0.5\_7000  & 2067188     & 2046886.5  &  & \textbf{2151921}   & 2141828.6 \\
bmcp\_5000\_5200\_0.3\_10000 & 2231359     & 2220887.2  &  & \textbf{2354408}   & 2345758.7 \\
bmcp\_5000\_5200\_0.5\_7000  & 2129556     & 2115338.3  &  & \textbf{2222169}   & 2211650.9 \\ \hline
Average                      & 1967745.0   & 1943665.3  &  & \textbf{2042027.0} & 2033356.7 \\ \bottomrule
\end{tabular}}
\end{table}

\begin{table}[t]
\centering
\caption{Comparison of the VDLS and its variant algorithms, VDLS$_{full}$ and VDLS$_{random}$, with 30 minutes of time limit in solving the 30 instances of the first set. Best results appear in bold.}
\label{table_v1}
\scalebox{0.65}{\begin{tabular}{lrrrrrrrr} \toprule
\multirow{2}{*}{instances}    & \multicolumn{2}{r}{VDLS$_{full}$} &  & \multicolumn{2}{r}{VDLS$_{random}$} &  & \multicolumn{2}{r}{VDLS}    \\ \cline{2-3} \cline{5-6} \cline{8-9} 
                              & best                & average     &  & best                & average       &  & best             & average  \\ \hline
bmcp\_585\_600\_0.05\_2000    & \textbf{71102}      & 71054.7     &  & 70752               & 70639.4       &  & \textbf{71102}   & 71034.4  \\
bmcp\_585\_600\_0.075\_1500   & \textbf{71025}      & 70671.5     &  & \textbf{71025}      & 70385.3       &  & \textbf{71025}   & 70492.5  \\
bmcp\_600\_585\_0.05\_2000    & \textbf{67636}      & 67636.0     &  & \textbf{67636}      & 67575.5       &  & \textbf{67636}   & 67621.6  \\
bmcp\_600\_585\_0.075\_1500   & \textbf{70588}      & 70353.0     &  & 70362               & 69900.9       &  & \textbf{70588}   & 70277.0  \\
bmcp\_600\_600\_0.05\_2000    & \textbf{68738}      & 68608.9     &  & 68387               & 68084.2       &  & \textbf{68738}   & 68580.7  \\
bmcp\_600\_600\_0.075\_1500   & \textbf{71904}      & 71780.1     &  & \textbf{71904}      & 71615.0       &  & \textbf{71904}   & 71780.7  \\
bmcp\_685\_700\_0.05\_2000    & \textbf{81227}      & 80538.2     &  & 79879               & 79852.2       &  & \textbf{81227}   & 80482.5  \\
bmcp\_685\_700\_0.075\_1500   & \textbf{83286}      & 82985.2     &  & 82931               & 82578.8       &  & \textbf{83286}   & 82886.6  \\
bmcp\_700\_685\_0.05\_2000    & \textbf{78054}      & 77751.9     &  & 77399               & 77066.5       &  & \textbf{78054}   & 77742.9  \\
bmcp\_700\_685\_0.075\_1500   & \textbf{78869}      & 78473.4     &  & 78232               & 77700.2       &  & \textbf{78869}   & 78642.9  \\
bmcp\_700\_700\_0.05\_2000    & \textbf{78458}      & 78272.6     &  & 78023               & 77488.9       &  & \textbf{78458}   & 78200.7  \\
bmcp\_700\_700\_0.075\_1500   & \textbf{84576}      & 84576.0     &  & \textbf{84576}      & 84204.3       &  & \textbf{84576}   & 84576.0  \\
bmcp\_785\_800\_0.05\_2000    & \textbf{92740}      & 92532.2     &  & \textbf{92740}      & 92141.5       &  & \textbf{92740}   & 92419.7  \\
bmcp\_785\_800\_0.075\_1500   & \textbf{95221}      & 95221.0     &  & \textbf{95221}      & 95116.5       &  & \textbf{95221}   & 95221.0  \\
bmcp\_800\_785\_0.05\_2000    & \textbf{89138}      & 89079.4     &  & 89084               & 88503.9       &  & \textbf{89138}   & 89054.7  \\
bmcp\_800\_785\_0.075\_1500   & \textbf{91856}      & 91723.8     &  & 91567               & 91006.4       &  & \textbf{91856}   & 91722.9  \\
bmcp\_800\_800\_0.05\_2000    & \textbf{91795}      & 91676.3     &  & 91443               & 90774.7       &  & \textbf{91795}   & 91604.5  \\
bmcp\_800\_800\_0.075\_1500   & \textbf{95995}      & 95590.8     &  & 95079               & 94903.0       &  & \textbf{95995}   & 95568.5  \\
bmcp\_885\_900\_0.05\_2000    & \textbf{102277}     & 102131.9    &  & 101398              & 100864.9      &  & \textbf{102277}  & 102059.2 \\
bmcp\_885\_900\_0.075\_1500   & \textbf{106940}     & 106529.1    &  & 106151              & 105876.7      &  & \textbf{106940}  & 106446.9 \\
bmcp\_900\_885\_0.05\_2000    & \textbf{99590}      & 99404.4     &  & 98680               & 98570.0       &  & \textbf{99590}   & 99368.7  \\
bmcp\_900\_885\_0.075\_1500   & \textbf{105141}     & 104991.6    &  & 104643              & 104248.2      &  & \textbf{105141}  & 105024.8 \\
bmcp\_900\_900\_0.05\_2000    & 101582              & 101185.0    &  & 100584              & 99787.1       &  & \textbf{102055}  & 101481.4 \\
bmcp\_900\_900\_0.075\_1500   & \textbf{105081}     & 104828.7    &  & 104211              & 103871.0      &  & \textbf{105081}  & 104919.1 \\
bmcp\_985\_1000\_0.05\_2000   & 110228              & 109700.8    &  & 109332              & 108917.7      &  & \textbf{110669}  & 109792.6 \\
bmcp\_985\_1000\_0.075\_1500  & \textbf{115505}     & 114977.6    &  & 114315              & 113877.7      &  & \textbf{115505}  & 114765.3 \\
bmcp\_1000\_985\_0.05\_2000   & \textbf{112057}     & 111768.8    &  & 111482              & 110461.2      &  & \textbf{112057}  & 111624.7 \\
bmcp\_1000\_985\_0.075\_1500  & \textbf{113615}     & 113153.5    &  & 112396              & 111992.6      &  & \textbf{113615}  & 113096.1 \\
bmcp\_1000\_1000\_0.05\_2000  & \textbf{113331}     & 113120.6    &  & 112893              & 112407.7      &  & \textbf{113331}  & 113129.6 \\
bmcp\_1000\_1000\_0.075\_1500 & \textbf{120246}     & 119954.1    &  & 119301              & 118994.4      &  & \textbf{120246}  & 119994.5 \\ \hline
Average                       & 92260.0             & 92009.0     &  & 91720.9             & 91313.5       &  & \textbf{92290.5} & 91987.1 
 \\ \bottomrule
\end{tabular}}
\end{table}

\begin{table}[!t]
\centering
\caption{Comparison of the VDLS and its variant algorithms, VDLS$_{full}$ and VDLS$_{random}$, with 30 minutes of time limit in solving the 30 instances of the second set. Best results appear in bold.}
\label{table_v2}
\scalebox{0.6}{\begin{tabular}{lrrrrrrrr} \toprule
\multirow{2}{*}{instances}  & \multicolumn{2}{r}{VDLS$_{full}$} &  & \multicolumn{2}{r}{VDLS$_{random}$} &  & \multicolumn{2}{r}{VDLS}     \\ \cline{2-3} \cline{5-6} \cline{8-9} 
                            & best                & average     &  & best             & average          &  & best              & average  \\ \hline
bmcp\_1100\_1000\_0.3\_3000 & \textbf{143475}     & 143048.1    &  & 141054           & 141054.0         &  & \textbf{143475}   & 143224.6 \\
bmcp\_1100\_1000\_0.5\_2000 & \textbf{127157}     & 126397.4    &  & 125223           & 125223.0         &  & 126917            & 126021.6 \\
bmcp\_1100\_1100\_0.3\_3000 & \textbf{157530}     & 156853.9    &  & 155669           & 155669.0         &  & \textbf{157530}   & 156462.5 \\
bmcp\_1100\_1100\_0.5\_2000 & \textbf{139141}     & 138861.0    &  & 135927           & 135220.5         &  & \textbf{139141}   & 138719.4 \\
bmcp\_1100\_1200\_0.3\_3000 & \textbf{172660}     & 171970.2    &  & 170981           & 169663.5         &  & \textbf{172660}   & 172075.3 \\
bmcp\_1100\_1200\_0.5\_2000 & \textbf{150493}     & 149742.0    &  & 147141           & 146614.1         &  & \textbf{150493}   & 149944.0 \\
bmcp\_1200\_1100\_0.3\_3000 & \textbf{161848}     & 161832.0    &  & 161640           & 160747.3         &  & \textbf{161848}   & 161848.0 \\
bmcp\_1200\_1100\_0.5\_2000 & \textbf{137700}     & 136515.4    &  & 135501           & 134691.7         &  & \textbf{137700}   & 136831.2 \\
bmcp\_1200\_1200\_0.3\_3000 & \textbf{172595}     & 172051.9    &  & 169986           & 169256.2         &  & \textbf{172595}   & 172151.8 \\
bmcp\_1200\_1200\_0.5\_2000 & \textbf{151440}     & 149296.0    &  & 148339           & 147277.5         &  & \textbf{151440}   & 149610.0 \\
bmcp\_1200\_1300\_0.3\_3000 & 184431              & 184000.4    &  & 182282           & 181571.8         &  & \textbf{186083}   & 183599.0 \\
bmcp\_1200\_1300\_0.5\_2000 & \textbf{162154}     & 161007.4    &  & 159028           & 157862.8         &  & \textbf{162154}   & 161146.9 \\
bmcp\_1300\_1200\_0.3\_3000 & 170372              & 169693.6    &  & 167575           & 167277.2         &  & \textbf{170786}   & 169724.7 \\
bmcp\_1300\_1200\_0.5\_2000 & \textbf{151402}     & 150846.1    &  & 149747           & 148738.3         &  & \textbf{151402}   & 151049.4 \\
bmcp\_1300\_1300\_0.3\_3000 & \textbf{184699}     & 183352.8    &  & 182406           & 181178.4         &  & \textbf{184699}   & 183239.0 \\
bmcp\_1300\_1300\_0.5\_2000 & \textbf{162146}     & 161351.7    &  & 160361           & 159282.4         &  & \textbf{162146}   & 161536.7 \\
bmcp\_1300\_1400\_0.3\_3000 & \textbf{199134}     & 198864.8    &  & 195811           & 194480.4         &  & \textbf{199134}   & 198889.4 \\
bmcp\_1300\_1400\_0.5\_2000 & \textbf{173954}     & 173797.5    &  & 172420           & 171367.5         &  & \textbf{173954}   & 173799.2 \\
bmcp\_1400\_1300\_0.3\_3000 & \textbf{183987}     & 183838.8    &  & 182685           & 181441.1         &  & \textbf{183987}   & 183796.3 \\
bmcp\_1400\_1300\_0.5\_2000 & \textbf{162544}     & 162407.9    &  & 160894           & 160422.0         &  & \textbf{162544}   & 162312.9 \\
bmcp\_1400\_1400\_0.3\_3000 & \textbf{200136}     & 199564.8    &  & 198789           & 196983.3         &  & \textbf{200136}   & 199616.7 \\
bmcp\_1400\_1400\_0.5\_2000 & 174737              & 173783.8    &  & 171921           & 170531.9         &  & \textbf{175583}   & 173984.9 \\
bmcp\_1400\_1500\_0.3\_3000 & 212221              & 211189.7    &  & 211156           & 208591.6         &  & \textbf{212492}   & 211264.4 \\
bmcp\_1400\_1500\_0.5\_2000 & \textbf{186818}     & 186332.1    &  & 185133           & 184300.0         &  & \textbf{186818}   & 186250.3 \\
bmcp\_1500\_1400\_0.3\_3000 & 197184              & 196639.3    &  & 195348           & 194293.1         &  & \textbf{197219}   & 196703.4 \\
bmcp\_1500\_1400\_0.5\_2000 & \textbf{174370}     & 173451.4    &  & 171763           & 171206.7         &  & 174284            & 173576.4 \\
bmcp\_1500\_1500\_0.3\_3000 & \textbf{211108}     & 210512.8    &  & 209275           & 207887.5         &  & \textbf{211108}   & 210393.7 \\
bmcp\_1500\_1500\_0.5\_2000 & \textbf{187052}     & 185793.0    &  & 185015           & 183905.1         &  & \textbf{187052}   & 185221.6 \\
bmcp\_1500\_1600\_0.3\_3000 & \textbf{226408}     & 224819.5    &  & 223377           & 222280.0         &  & \textbf{226408}   & 224926.6 \\
bmcp\_1500\_1600\_0.5\_2000 & 198672              & 198259.5    &  & 196923           & 196291.1         &  & \textbf{199678}   & 198462.6 \\ \hline
Average                     & 173918.9            & 173202.5    &  & 171779.0         & 170843.6         &  & \textbf{174048.9} & 173212.8
 \\ \bottomrule \vspace{-2em}
\end{tabular}}
\end{table}

\begin{table}[t]
\centering
\caption{Comparison of the VDLS and its variant algorithms, VDLS$_{full}$ and VDLS$_{random}$, with 30 minutes of time limit in solving the 30 instances of the third set. Best results appear in bold.}
\label{table_v3}
\scalebox{0.56}{\begin{tabular}{lrrrrrrrr} \toprule
\multirow{2}{*}{instances}   & \multicolumn{2}{r}{VDLS$_{full}$} &  & \multicolumn{2}{r}{VDLS$_{random}$} &  & \multicolumn{2}{r}{VDLS}       \\ \cline{2-3} \cline{5-6} \cline{8-9} 
                             & best                & average     &  & best             & average          &  & best               & average   \\ \hline
bmcp\_4200\_4000\_0.3\_10000 & \textbf{1835947}    & 1829767.7   &  & 1799099          & 1793354.4        &  & \textbf{1835947}   & 1833108.2 \\
bmcp\_4200\_4000\_0.5\_7000  & 1733215             & 1725738.5   &  & 1701194          & 1696877.0        &  & \textbf{1738553}   & 1729188.0 \\
bmcp\_4200\_4200\_0.3\_10000 & 1923952             & 1918736.4   &  & 1892754          & 1890927.2        &  & \textbf{1925706}   & 1918624.2 \\
bmcp\_4200\_4200\_0.5\_7000  & 1809058             & 1803212.4   &  & 1779127          & 1777656.9        &  & \textbf{1811649}   & 1801880.2 \\
bmcp\_4200\_4400\_0.3\_10000 & 2022834             & 2020204.0   &  & 2002056          & 1997189.7        &  & \textbf{2023093}   & 2020209.8 \\
bmcp\_4200\_4400\_0.5\_7000  & 1897776             & 1888715.3   &  & 1870362          & 1863878.0        &  & \textbf{1903515}   & 1889809.5 \\
bmcp\_4400\_4200\_0.3\_10000 & 1917530             & 1908724.7   &  & 1890371          & 1885678.4        &  & \textbf{1920418}   & 1911396.2 \\
bmcp\_4400\_4200\_0.5\_7000  & 1808313             & 1804231.2   &  & 1791317          & 1788369.0        &  & \textbf{1810031}   & 1805200.7 \\
bmcp\_4400\_4400\_0.3\_10000 & \textbf{2027801}    & 2022493.1   &  & 1998685          & 1997515.4        &  & \textbf{2027801}   & 2024036.1 \\
bmcp\_4400\_4400\_0.5\_7000  & 1890349             & 1884896.0   &  & 1849547          & 1847377.2        &  & \textbf{1901941}   & 1888300.5 \\
bmcp\_4400\_4600\_0.3\_10000 & 2097111             & 2091973.3   &  & 2070880          & 2067927.9        &  & \textbf{2100925}   & 2093610.1 \\
bmcp\_4400\_4600\_0.5\_7000  & 1981566             & 1976728.2   &  & 1968883          & 1966526.8        &  & \textbf{1989582}   & 1979416.6 \\
bmcp\_4600\_4400\_0.3\_10000 & 2007323             & 2002213.2   &  & 1975158          & 1969241.9        &  & \textbf{2008258}   & 2001368.8 \\
bmcp\_4600\_4400\_0.5\_7000  & 1889059             & 1886725.1   &  & 1880145          & 1880145.0        &  & \textbf{1898602}   & 1889658.5 \\
bmcp\_4600\_4600\_0.3\_10000 & 2103993             & 2097677.7   &  & 2080601          & 2065594.8        &  & \textbf{2106938}   & 2100124.5 \\
bmcp\_4600\_4600\_0.5\_7000  & 1978206             & 1972638.7   &  & 1953721          & 1953251.9        &  & \textbf{1980462}   & 1973677.2 \\
bmcp\_4600\_4800\_0.3\_10000 & \textbf{2201977}    & 2197678.1   &  & 2171744          & 2171497.4        &  & \textbf{2201977}   & 2197572.4 \\
bmcp\_4600\_4800\_0.5\_7000  & 2069048             & 2058651.2   &  & 2039254          & 2036251.6        &  & \textbf{2070913}   & 2057914.3 \\
bmcp\_4800\_4600\_0.3\_10000 & 2091456             & 2084178.6   &  & 2055984          & 2055318.2        &  & \textbf{2096747}   & 2087482.0 \\
bmcp\_4800\_4600\_0.5\_7000  & 1980482             & 1971733.9   &  & 1937401          & 1937401.0        &  & \textbf{1991409}   & 1973721.9 \\
bmcp\_4800\_4800\_0.3\_10000 & 2181967             & 2171730.1   &  & 2151323          & 2143758.3        &  & \textbf{2183525}   & 2175464.8 \\
bmcp\_4800\_4800\_0.5\_7000  & 2064768             & 2058002.9   &  & 2029868          & 2029556.6        &  & \textbf{2066487}   & 2059458.7 \\
bmcp\_4800\_5000\_0.3\_10000 & 2278226             & 2273093.1   &  & 2261424          & 2261424.0        &  & \textbf{2280317}   & 2273869.0 \\
bmcp\_4800\_5000\_0.5\_7000  & 2140951             & 2133532.0   &  & 2110025          & 2110025.0        &  & \textbf{2146413}   & 2136313.2 \\
bmcp\_5000\_4800\_0.3\_10000 & 2183270             & 2174154.3   &  & 2144805          & 2135011.9        &  & \textbf{2183483}   & 2175171.2 \\
bmcp\_5000\_4800\_0.5\_7000  & 2064436             & 2053364.9   &  & 2036595          & 2036595.0        &  & \textbf{2066000}   & 2053359.9 \\
bmcp\_5000\_5000\_0.3\_10000 & 2257921             & 2250425.2   &  & 2232094          & 2222644.0        &  & \textbf{2261620}   & 2251527.0 \\
bmcp\_5000\_5000\_0.5\_7000  & 2145398             & 2140167.1   &  & 2119843          & 2119843.0        &  & \textbf{2151921}   & 2141828.6 \\
bmcp\_5000\_5200\_0.3\_10000 & 2352860             & 2345247.5   &  & 2313592          & 2310718.3        &  & \textbf{2354408}   & 2345758.7 \\
bmcp\_5000\_5200\_0.5\_7000  & 2214566             & 2208215.9   &  & 2193299          & 2191433.5        &  & \textbf{2222169}   & 2211650.9 \\ \hline
Average                      & 2038378.6           & 2031828.3   &  & 2010038.4        & 2006766.3        &  & \textbf{2042027.0} & 2033356.7 \\ \bottomrule \vspace{-2.5em}
\end{tabular}}
\end{table}

\subsection{Parameter Study}
To analyze the effectiveness of the parameters including $d$ and $k$ on the performance of VDLS, and determine the best setting of them, we compare VDLS with $d \in [5, 10]$ and $k \in [5, 10]$ in solving the 30 instances of the first set, under 10 or 30 minutes of time limit. The results are shown in Table \ref{table_para}. The results are expressed by the average value of the best solutions of all the 30 instances obtained by the algorithms in 10 runs. From the results we can observe that:

(1) VDLS with medium values of $d$ and $k$ shows better performance. For example, $d = 8$, $k = 5,6,7$ and $d = 7$, $k = 8$. Actually, the larger the value of $d$ or $k$, the higher the quality of the local optima for VDLS, and the lower the algorithm efficiency. Therefore, a medium value of $d$ or $k$ can well trade-off the search ability and the algorithm efficiency.

(2) The proposed VDLS algorithm is robust and effective, since VDLS with any $d \in [5,10]$ and $k \in [5,10]$ outperforms the PLTS algorithm \cite{Li2021}, which yields a result of 91533.7 with 10 minutes of time limit and 91733.9 with 30 minutes of time limit (see detailed results in Tables \ref{table1-10} and \ref{table1-30}).

According to the results in Table \ref{table_para}, we finally select $d = 8$ and $k = 7$ as the default parameters of the VDLS algorithm. The results of VDLS in the rest of the paper are obtained by VDLS with the default parameters.

\subsection{Comparison on the VDLS and the Baselines}
This subsection presents the comparison results of the proposed VDLS algorithm and the baseline algorithms, including the CPLEX solver and the PLTS. We compare the best solutions, the average solutions of the VDLS and the PLTS for each instance in 10 runs, as well as the lower bound (LB) and upper bound (UB) of the CPLEX solver. 

The results of the CPLEX solver with 2 hours and 5 hours of time limit, the PLTS and the VDLS with 10 minutes and 30 minutes of time limit, in solving the first set of instances are shown in Tables~\ref{table1-10} and~\ref{table1-30}, respectively. The results of the PLTS and the VDLS algorithms with 10 or 30 minutes of time limit in solving the second and third sets of instances are shown in Tables~\ref{table2} and~\ref{table3}, respectively. 

From the results, we can observe that:

(1) The VDLS outperforms the CPLEX solver in either 2 hours or 5 hours of time limit in solving all the 30 instances of the first set. And with either 10 minutes or 30 minutes of time limit, the VDLS significantly outperforms the PLTS algorithm in solving the three sets of instances, indicating the excellent performance of the proposed VDLS algorithm for the BMCP.

(2) The advantages of the VDLS over the PLTS are more obvious when solving the large scale instances belonging to the second and third sets, which have complex structural features, indicating that the proposed VDLS algorithm is more effective for solving large BMCP instances than the PLTS, and the proposed neighbour structure is effective for solving the BMCP instances with diverse structures. The results also indicate that our designed 60 instances can distinguish the performance of different algorithms better.

\subsection{Ablation Study on the Branching Strategy}
The branching strategy in the tree search process of the VDLS algorithm directly influences the efficiency of the local search process and the quality of the solution. The branching strategy in the VDLS selects at most $k$ items with the largest \textit{gain} value (i.e., Eq.~\ref{eq_gain}) in the candidate set $U$ (line 8 in Algorithm~\ref{alg_LS}) that consists of all the neighbours of the last flipping item that have not been visited during this local search process and flipping each of them will result in a feasible solution. 

In order to analyze the performance of the neighbour structure defined in the VDLS as well as the greedy approach when selecting the branch nodes, we compare the VDLS with two variant algorithms. 
\begin{itemize}
    \item The first variant algorithm $\mathrm{VDLS}_{full}$ allows the algorithm to select the items not belonging to the neighbours of the last flipping item as the branch nodes, (i.e., replace $\Gamma(i)$ with $I$ to generate $U$). 
    \item The second variant algorithm $\mathrm{VDLS}_{random}$ randomly selects at most $k$ items in $U$.
\end{itemize}
Both variant algorithms run 10 times with 30 minutes of time limit for each instance of the three sets. The comparison results on the three sets of instances are shown in Tables~\ref{table_v1},~\ref{table_v2} and~\ref{table_v3}, respectively. The results are expressed by the best solution and average solution in 10 runs of each instance.

From the results in Tables~\ref{table_v1},~\ref{table_v2} and~\ref{table_v3} we can observe that, the performance of the VDLS is better than the $\mathrm{VDLS}_{full}$ in solving the three sets of instances, demonstrating that the neighbour structure we defined in the VDLS is reliable, and using neighbours as branch candidate set instead of all the items $I$ can improve the efficiency of the algorithm by abandoning relatively low-quality branch nodes. Moreover, the performance of the VDLS is also better than the $\mathrm{VDLS}_{random}$, indicating that the greedy approach used in the branching strategy of the VDLS is effective.

\begin{table}[t]
\centering
\caption{Comparison of the VDLS and its variant algorithms, VDLS$_{empty\_init}$ and VDLS$_{random\_init}$, as well as the PLTS algorithm, with 30 minutes of time limit in solving the 30 instances of the first set. Best results appear in bold.}
\label{table_v4}
\scalebox{0.6}{\begin{tabular}{lrrrrrrrrrrr} \toprule
\multirow{2}{*}{Instances}    & \multicolumn{2}{r}{PLTS}   &  & \multicolumn{2}{r}{VDLS$_{empty\_init}$} &  & \multicolumn{2}{r}{VDLS$_{random\_init}$} &  & \multicolumn{2}{r}{VDLS}    \\ \cline{2-3} \cline{5-6} \cline{8-9} \cline{11-12} 
                              & best            & average  &  & best                   & average         &  & best                    & average         &  & best             & average  \\ \hline
bmcp\_585\_600\_0.05\_2000    & \textbf{71102}  & 70915.0  &  & \textbf{71102}         & 71013.3         &  & \textbf{71102}          & 71021.2         &  & \textbf{71102}   & 71034.4  \\
bmcp\_585\_600\_0.075\_1500   & 70677           & 70670.2  &  & \textbf{71025}         & 70472.0         &  & \textbf{71025}          & 70455.0         &  & \textbf{71025}   & 70492.5  \\
bmcp\_600\_585\_0.05\_2000    & \textbf{67636}  & 67041.5  &  & \textbf{67636}         & 67588.8         &  & \textbf{67636}          & 67621.6         &  & \textbf{67636}   & 67621.6  \\
bmcp\_600\_585\_0.075\_1500   & 70318           & 70318.0  &  & \textbf{70588}         & 70155.1         &  & \textbf{70588}          & 70192.9         &  & \textbf{70588}   & 70277.0  \\
bmcp\_600\_600\_0.05\_2000    & 68453           & 68386.8  &  & \textbf{68738}         & 68521.4         &  & 68634                   & 68512.1         &  & \textbf{68738}   & 68580.7  \\
bmcp\_600\_600\_0.075\_1500   & 71746           & 71746.0  &  & \textbf{71904}         & 71720.3         &  & \textbf{71904}          & 71730.3         &  & \textbf{71904}   & 71780.7  \\
bmcp\_685\_700\_0.05\_2000    & 80197           & 80096.6  &  & \textbf{81227}         & 80934.6         &  & \textbf{81227}          & 81009.0         &  & \textbf{81227}   & 80482.5  \\
bmcp\_685\_700\_0.075\_1500   & 82955           & 82944.2  &  & 82955                  & 82823.8         &  & \textbf{83286}          & 82840.5         &  & \textbf{83286}   & 82886.6  \\
bmcp\_700\_685\_0.05\_2000    & 77876           & 77372.7  &  & \textbf{78054}         & 77915.0         &  & \textbf{78054}          & 77760.8         &  & \textbf{78054}   & 77742.9  \\
bmcp\_700\_685\_0.075\_1500   & \textbf{78869}  & 78810.4  &  & \textbf{78869}         & 78759.3         &  & \textbf{78869}          & 78547.3         &  & \textbf{78869}   & 78642.9  \\
bmcp\_700\_700\_0.05\_2000    & 78028           & 77501.5  &  & \textbf{78458}         & 78181.1         &  & \textbf{78458}          & 78238.8         &  & \textbf{78458}   & 78200.7  \\
bmcp\_700\_700\_0.075\_1500   & 84570           & 83720.7  &  & \textbf{84576}         & 84576.0         &  & \textbf{84576}          & 84576.0         &  & \textbf{84576}   & 84576.0  \\
bmcp\_785\_800\_0.05\_2000    & 92431           & 92195.7  &  & 92608                  & 92477.3         &  & 92608                   & 92339.7         &  & \textbf{92740}   & 92419.7  \\
bmcp\_785\_800\_0.075\_1500   & 94245           & 94245.0  &  & \textbf{95221}         & 94466.1         &  & \textbf{95221}          & 94641.6         &  & \textbf{95221}   & 95221.0  \\
bmcp\_800\_785\_0.05\_2000    & 88562           & 88560.3  &  & \textbf{89138}         & 88911.4         &  & \textbf{89138}          & 89049.4         &  & \textbf{89138}   & 89054.7  \\
bmcp\_800\_785\_0.075\_1500   & 91021           & 90971.8  &  & \textbf{91856}         & 91435.5         &  & \textbf{91856}          & 91320.7         &  & \textbf{91856}   & 91722.9  \\
bmcp\_800\_800\_0.05\_2000    & 91748           & 90934.2  &  & \textbf{91795}         & 91686.4         &  & \textbf{91795}          & 91446.4         &  & \textbf{91795}   & 91604.5  \\
bmcp\_800\_800\_0.075\_1500   & 95533           & 95533.0  &  & 95746                  & 95298.2         &  & \textbf{95995}          & 95340.6         &  & \textbf{95995}   & 95568.5  \\
bmcp\_885\_900\_0.05\_2000    & 101293          & 101142.0 &  & \textbf{102277}        & 102031.0        &  & \textbf{102277}         & 102049.7        &  & \textbf{102277}  & 102059.2 \\
bmcp\_885\_900\_0.075\_1500   & 106364          & 105782.6 &  & \textbf{106940}        & 106719.6        &  & \textbf{106940}         & 106618.8        &  & \textbf{106940}  & 106446.9 \\
bmcp\_900\_885\_0.05\_2000    & 98840           & 98645.4  &  & \textbf{99590}         & 99203.7         &  & \textbf{99590}          & 99304.8         &  & \textbf{99590}   & 99368.7  \\
bmcp\_900\_885\_0.075\_1500   & \textbf{105141} & 103846.9 &  & \textbf{105141}        & 105091.2        &  & \textbf{105141}         & 105041.4        &  & \textbf{105141}  & 105024.8 \\
bmcp\_900\_900\_0.05\_2000    & 101265          & 101211.4 &  & \textbf{102055}        & 101366.5        &  & \textbf{102055}         & 101467.8        &  & \textbf{102055}  & 101481.4 \\
bmcp\_900\_900\_0.075\_1500   & 104521          & 104521.0 &  & \textbf{105081}        & 104642.9        &  & \textbf{105081}         & 104629.0        &  & \textbf{105081}  & 104919.1 \\
bmcp\_985\_1000\_0.05\_2000   & 109567          & 108938.0 &  & \textbf{110669}        & 109936.4        &  & \textbf{110669}         & 110095.6        &  & \textbf{110669}  & 109792.6 \\
bmcp\_985\_1000\_0.075\_1500  & 113541          & 113541.0 &  & \textbf{115505}        & 114943.2        &  & \textbf{115505}         & 115188.6        &  & \textbf{115505}  & 114765.3 \\
bmcp\_1000\_985\_0.05\_2000   & 111176          & 110712.6 &  & \textbf{112057}        & 111854.8        &  & \textbf{112057}         & 111570.4        &  & \textbf{112057}  & 111624.7 \\
bmcp\_1000\_985\_0.075\_1500  & 112250          & 112218.9 &  & \textbf{113615}        & 113021.8        &  & \textbf{113615}         & 112743.9        &  & \textbf{113615}  & 113096.1 \\
bmcp\_1000\_1000\_0.05\_2000  & 112640          & 111577.8 &  & \textbf{113331}        & 113165.4        &  & \textbf{113331}         & 113140.3        &  & \textbf{113331}  & 113129.6 \\
bmcp\_1000\_1000\_0.075\_1500 & 119453          & 118499.3 &  & \textbf{120246}        & 119660.3        &  & \textbf{120246}         & 119913.0        &  & \textbf{120246}  & 119994.5 \\ \hline
Average                       & 91733.9         & 91420.0  &  & 92266.8                & 91952.4         &  & 92282.6                 & 91946.9         &  & \textbf{92290.5} & 91987.1 
 \\ \bottomrule
\end{tabular}}
\end{table}

\begin{table}[!t]
\centering
\caption{Comparison of the VDLS and its variant algorithms, VDLS$_{empty\_init}$ and VDLS$_{random\_init}$, as well as the PLTS algorithm, with 30 minutes of time limit in solving the 30 instances of the second set. Best results appear in bold.}
\label{table_v5}
\scalebox{0.6}{\begin{tabular}{lrrrrrrrrrrr} \toprule
\multirow{2}{*}{Instances}  & \multicolumn{2}{r}{PLTS}   &  & \multicolumn{2}{r}{VDLS$_{empty\_init}$} &  & \multicolumn{2}{r}{VDLS$_{random\_init}$} &  & \multicolumn{2}{r}{VDLS}     \\ \cline{2-3} \cline{5-6} \cline{8-9} \cline{11-12} 
                            & best            & average  &  & best                   & average         &  & best                    & average         &  & best              & average  \\ \hline
bmcp\_1100\_1000\_0.3\_3000 & 142835          & 142428.7 &  & \textbf{143475}        & 143052.0        &  & \textbf{143475}         & 143317.0        &  & \textbf{143475}   & 143224.6 \\
bmcp\_1100\_1000\_0.5\_2000 & 125974          & 125790.0 &  & \textbf{127157}        & 126628.5        &  & 126792                  & 126400.0        &  & 126917            & 126021.6 \\
bmcp\_1100\_1100\_0.3\_3000 & 154936          & 154661.7 &  & \textbf{157530}        & 156913.1        &  & 157264                  & 156684.0        &  & \textbf{157530}   & 156462.5 \\
bmcp\_1100\_1100\_0.5\_2000 & \textbf{139196} & 138169.0 &  & \textbf{139196}        & 138470.8        &  & \textbf{139196}         & 138679.7        &  & 139141            & 138719.4 \\
bmcp\_1100\_1200\_0.3\_3000 & 169113          & 168763.6 &  & \textbf{172660}        & 172348.5        &  & \textbf{172660}         & 172137.0        &  & \textbf{172660}   & 172075.3 \\
bmcp\_1100\_1200\_0.5\_2000 & 150301          & 149366.4 &  & \textbf{150493}        & 149920.9        &  & \textbf{150493}         & 150039.2        &  & \textbf{150493}   & 149944.0 \\
bmcp\_1200\_1100\_0.3\_3000 & \textbf{161848} & 159620.3 &  & \textbf{161848}        & 161848.0        &  & \textbf{161848}         & 161848.0        &  & \textbf{161848}   & 161848.0 \\
bmcp\_1200\_1100\_0.5\_2000 & 135858          & 135509.2 &  & 137046                 & 136658.3        &  & 136995                  & 136380.8        &  & \textbf{137700}   & 136831.2 \\
bmcp\_1200\_1200\_0.3\_3000 & 169219          & 168821.7 &  & \textbf{172595}        & 172238.3        &  & \textbf{172595}         & 172154.5        &  & \textbf{172595}   & 172151.8 \\
bmcp\_1200\_1200\_0.5\_2000 & 149654          & 148679.1 &  & \textbf{151440}        & 149839.0        &  & \textbf{151440}         & 149991.3        &  & \textbf{151440}   & 149610.0 \\
bmcp\_1200\_1300\_0.3\_3000 & 183004          & 182220.4 &  & 184384                 & 183829.5        &  & 184195                  & 183609.9        &  & \textbf{186083}   & 183599.0 \\
bmcp\_1200\_1300\_0.5\_2000 & 160369          & 159735.4 &  & \textbf{162154}        & 161180.2        &  & \textbf{162154}         & 161188.3        &  & \textbf{162154}   & 161146.9 \\
bmcp\_1300\_1200\_0.3\_3000 & 168586          & 168473.8 &  & 170283                 & 169933.0        &  & \textbf{170786}         & 169874.1        &  & \textbf{170786}   & 169724.7 \\
bmcp\_1300\_1200\_0.5\_2000 & 151067          & 150034.7 &  & 151402                 & 150650.5        &  & \textbf{151512}         & 150982.9        &  & 151402            & 151049.4 \\
bmcp\_1300\_1300\_0.3\_3000 & 180608          & 179978.4 &  & \textbf{184699}        & 183104.5        &  & 184627                  & 183237.6        &  & \textbf{184699}   & 183239.0 \\
bmcp\_1300\_1300\_0.5\_2000 & 160485          & 159511.3 &  & \textbf{162146}        & 161168.5        &  & \textbf{162146}         & 160950.0        &  & \textbf{162146}   & 161536.7 \\
bmcp\_1300\_1400\_0.3\_3000 & 197709          & 195681.9 &  & \textbf{199134}        & 198992.4        &  & \textbf{199134}         & 198914.6        &  & \textbf{199134}   & 198889.4 \\
bmcp\_1300\_1400\_0.5\_2000 & 173349          & 171665.0 &  & \textbf{173954}        & 173445.6        &  & \textbf{173954}         & 173559.3        &  & \textbf{173954}   & 173799.2 \\
bmcp\_1400\_1300\_0.3\_3000 & 182881          & 180143.5 &  & \textbf{183987}        & 183465.6        &  & 183914                  & 183677.0        &  & \textbf{183987}   & 183796.3 \\
bmcp\_1400\_1300\_0.5\_2000 & 161995          & 160647.0 &  & 162446                 & 161748.9        &  & 162465                  & 161986.3        &  & \textbf{162544}   & 162312.9 \\
bmcp\_1400\_1400\_0.3\_3000 & 199879          & 197229.0 &  & \textbf{200136}        & 199469.9        &  & \textbf{200136}         & 199824.5        &  & \textbf{200136}   & 199616.7 \\
bmcp\_1400\_1400\_0.5\_2000 & 174683          & 173393.1 &  & \textbf{175583}        & 174009.4        &  & 174683                  & 173992.4        &  & \textbf{175583}   & 173984.9 \\
bmcp\_1400\_1500\_0.3\_3000 & 210077          & 208315.6 &  & 212267                 & 211120.3        &  & 211494                  & 211274.1        &  & \textbf{212492}   & 211264.4 \\
bmcp\_1400\_1500\_0.5\_2000 & 184259          & 183289.9 &  & 185902                 & 185071.3        &  & 186619                  & 184816.6        &  & \textbf{186818}   & 186250.3 \\
bmcp\_1500\_1400\_0.3\_3000 & 194156          & 193554.8 &  & \textbf{197219}        & 196694.9        &  & \textbf{197219}         & 196650.0        &  & \textbf{197219}   & 196703.4 \\
bmcp\_1500\_1400\_0.5\_2000 & 173475          & 171426.9 &  & \textbf{174284}        & 173810.6        &  & \textbf{174284}         & 173953.5        &  & \textbf{174284}   & 173576.4 \\
bmcp\_1500\_1500\_0.3\_3000 & 209929          & 207795.3 &  & \textbf{211108}        & 210447.6        &  & \textbf{211108}         & 210857.9        &  & \textbf{211108}   & 210393.7 \\
bmcp\_1500\_1500\_0.5\_2000 & 184082          & 182676.2 &  & \textbf{187052}        & 185348.1        &  & 185931                  & 185251.3        &  & \textbf{187052}   & 185221.6 \\
bmcp\_1500\_1600\_0.3\_3000 & \textbf{226408} & 221641.3 &  & \textbf{226408}        & 224705.0        &  & \textbf{226408}         & 225218.9        &  & \textbf{226408}   & 224926.6 \\
bmcp\_1500\_1600\_0.5\_2000 & 198214          & 196659.0 &  & 198672                 & 198052.5        &  & 198834                  & 197922.9        &  & \textbf{199678}   & 198462.6 \\ \hline
Average                     & 172471.6        & 171196.1 &  & 173888.7               & 173138.9        &  & 173812.0                & 173179.1        &  & \textbf{174048.9} & 173212.8
 \\ \bottomrule \vspace{-2em}
\end{tabular}}
\end{table}

\begin{table}[t]
\centering
\caption{Comparison of the VDLS and its variant algorithms, VDLS$_{empty\_init}$ and VDLS$_{random\_init}$, as well as the PLTS algorithm, with 30 minutes of time limit in solving the 30 instances of the third set. Best results appear in bold.}
\label{table_v6}
\scalebox{0.56}{\begin{tabular}{lrrrrrrrrrrr} \toprule
\multirow{2}{*}{Instances}   & \multicolumn{2}{r}{PLTS} &  & \multicolumn{2}{r}{VDLS$_{empty\_init}$} &  & \multicolumn{2}{r}{VDLS$_{random\_init}$} &  & \multicolumn{2}{r}{VDLS}       \\ \cline{2-3} \cline{5-6} \cline{8-9} \cline{11-12} 
                             & best        & average    &  & best                   & average         &  & best                & average             &  & best               & average   \\ \hline
bmcp\_4200\_4000\_0.3\_10000 & 1778013     & 1747411.9  &  & 1834284                & 1803563.4       &  & 1833972             & 1803414.7           &  & \textbf{1835947}   & 1833108.2 \\
bmcp\_4200\_4000\_0.5\_7000  & 1685141     & 1666673.3  &  & 1731239                & 1711775.5       &  & 1728424             & 1706506.7           &  & \textbf{1738553}   & 1729188.0 \\
bmcp\_4200\_4200\_0.3\_10000 & 1874190     & 1839780.1  &  & 1891453                & 1880097.5       &  & 1906804             & 1885572.2           &  & \textbf{1925706}   & 1918624.2 \\
bmcp\_4200\_4200\_0.5\_7000  & 1764277     & 1742121.7  &  & 1799917                & 1774357.8       &  & 1807742             & 1779957.7           &  & \textbf{1811649}   & 1801880.2 \\
bmcp\_4200\_4400\_0.3\_10000 & 1954069     & 1932251.0  &  & 1978774                & 1971898.8       &  & 1976199             & 1972521.3           &  & \textbf{2023093}   & 2020209.8 \\
bmcp\_4200\_4400\_0.5\_7000  & 1853677     & 1820444.4  &  & 1888408                & 1861994.5       &  & 1901803             & 1860441.4           &  & \textbf{1903515}   & 1889809.5 \\
bmcp\_4400\_4200\_0.3\_10000 & 1833465     & 1816879.1  &  & 1915183                & 1878661.3       &  & 1885127             & 1878187.1           &  & \textbf{1920418}   & 1911396.2 \\
bmcp\_4400\_4200\_0.5\_7000  & 1746064     & 1730783.9  &  & 1795133                & 1771581.4       &  & 1777184             & 1766877.8           &  & \textbf{1810031}   & 1805200.7 \\
bmcp\_4400\_4400\_0.3\_10000 & 1945218     & 1922134.2  &  & 2022238                & 1995933.5       &  & 1985214             & 1978845.9           &  & \textbf{2027801}   & 2024036.1 \\
bmcp\_4400\_4400\_0.5\_7000  & 1844031     & 1821467.8  &  & 1895699                & 1864656.6       &  & 1865846             & 1853948.8           &  & \textbf{1901941}   & 1888300.5 \\
bmcp\_4400\_4600\_0.3\_10000 & 2003959     & 1993946.2  &  & 2089841                & 2048418.3       &  & 2098311             & 2057479.1           &  & \textbf{2100925}   & 2093610.1 \\
bmcp\_4400\_4600\_0.5\_7000  & 1937066     & 1906956.1  &  & 1978452                & 1947386.4       &  & 1988053             & 1943530.0           &  & \textbf{1989582}   & 1979416.6 \\
bmcp\_4600\_4400\_0.3\_10000 & 1930498     & 1913501.0  &  & 1971422                & 1963026.3       &  & 1995420             & 1966461.0           &  & \textbf{2008258}   & 2001368.8 \\
bmcp\_4600\_4400\_0.5\_7000  & 1845046     & 1817697.2  &  & 1889987                & 1857307.9       &  & 1887876             & 1863836.3           &  & \textbf{1898602}   & 1889658.5 \\
bmcp\_4600\_4600\_0.3\_10000 & 2027132     & 2005096.7  &  & 2103372                & 2059542.7       &  & 2057082             & 2048803.4           &  & \textbf{2106938}   & 2100124.5 \\
bmcp\_4600\_4600\_0.5\_7000  & 1921261     & 1900376.3  &  & 1971064                & 1937988.1       &  & 1950337             & 1938807.7           &  & \textbf{1980462}   & 1973677.2 \\
bmcp\_4600\_4800\_0.3\_10000 & 2120120     & 2091722.5  &  & 2201518                & 2177117.9       &  & 2196065             & 2168175.1           &  & \textbf{2201977}   & 2197572.4 \\
bmcp\_4600\_4800\_0.5\_7000  & 1989454     & 1967943.3  &  & 2062229                & 2019898.7       &  & 2065687             & 2040861.3           &  & \textbf{2070913}   & 2057914.3 \\
bmcp\_4800\_4600\_0.3\_10000 & 2019033     & 1991673.5  &  & 2054553                & 2044674.6       &  & 2085908             & 2049130.6           &  & \textbf{2096747}   & 2087482.0 \\
bmcp\_4800\_4600\_0.5\_7000  & 1922466     & 1901157.6  &  & 1981511                & 1943407.6       &  & 1971668             & 1942246.4           &  & \textbf{1991409}   & 1973721.9 \\
bmcp\_4800\_4800\_0.3\_10000 & 2100773     & 2058921.6  &  & 2176277                & 2130880.7       &  & 2172264             & 2130106.7           &  & \textbf{2183525}   & 2175464.8 \\
bmcp\_4800\_4800\_0.5\_7000  & 2009129     & 1973993.6  &  & \textbf{2068419}       & 2038013.3       &  & 2049709             & 2022473.7           &  & 2066487            & 2059458.7 \\
bmcp\_4800\_5000\_0.3\_10000 & 2184445     & 2157793.0  &  & 2224114                & 2216495.4       &  & 2268312             & 2226567.3           &  & \textbf{2280317}   & 2273869.0 \\
bmcp\_4800\_5000\_0.5\_7000  & 2087596     & 2050741.0  &  & 2133028                & 2092000.2       &  & 2134855             & 2103473.5           &  & \textbf{2146413}   & 2136313.2 \\
bmcp\_5000\_4800\_0.3\_10000 & 2069979     & 2051607.2  &  & 2158927                & 2121501.3       &  & 2168454             & 2127542.6           &  & \textbf{2183483}   & 2175171.2 \\
bmcp\_5000\_4800\_0.5\_7000  & 2013921     & 1977310.7  &  & 2055668                & 2021525.7       &  & 2064677             & 2027452.6           &  & \textbf{2066000}   & 2053359.9 \\
bmcp\_5000\_5000\_0.3\_10000 & 2144223     & 2126462.9  &  & 2239557                & 2199620.3       &  & 2206885             & 2201582.7           &  & \textbf{2261620}   & 2251527.0 \\
bmcp\_5000\_5000\_0.5\_7000  & 2067188     & 2046886.5  &  & 2098393                & 2088820.7       &  & 2120476             & 2096355.4           &  & \textbf{2151921}   & 2141828.6 \\
bmcp\_5000\_5200\_0.3\_10000 & 2231359     & 2220887.2  &  & 2331638                & 2294991.9       &  & 2332974             & 2289507.8           &  & \textbf{2354408}   & 2345758.7 \\
bmcp\_5000\_5200\_0.5\_7000  & 2129556     & 2115338.3  &  & 2217338                & 2175089.1       &  & 2209861             & 2171932.5           &  & \textbf{2222169}   & 2211650.9 \\ \hline
Average                      & 1967745.0   & 1943665.3  &  & 2025321.2              & 1996407.6       &  & 2023106.3           & 1996753.3           &  & \textbf{2042027.0} & 2033356.7
 \\ \bottomrule \vspace{-2em}
\end{tabular}}
\end{table}

\subsection{Ablation Study on the Initialization Method}
In order to evaluate the effectiveness of the greedy constructive algorithm in the VDLS, we further do an ablation study on the initialization method of the VDLS algorithm by comparing the VDLS with two other variants.

\begin{itemize}
    \item The third variant algorithm $\mathrm{VDLS}_{empty\_init}$ sets the initial solution $S = \emptyset$.
    \item The fourth variant algorithm $\mathrm{VDLS}_{random\_init}$ generates the initial solution by first randomly adding an item per step, until the total cost of the added items exceeding the budget. Then the algorithm removes the last item added and yields a feasible initial solution. 
\end{itemize}

We compare the best solutions and average solutions in 10 runs of the VDLS and its two variants described above, as well as the PLTS algorithm, with 30 minutes of time limit, in solving the three sets of BMCP instances. The results of the three sets of instances are shown in Tables~\ref{table_v4}~\ref{table_v5} and~\ref{table_v6}, respectively. From the results we observe that:

(1) The results of the VDLS are better than the results of the $\mathrm{VDLS}_{empty\_init}$ and $\mathrm{VDLS}_{random\_init}$ algorithms, especially for the large instances of the third set. The results indicate that the proposed local search method can yield better solutions with high-quality initial solutions, and the greedy constructive algorithm can improve the VDLS algorithm by providing high-quality initial solutions, especially for large instances.

(2) The $\mathrm{VDLS}_{empty\_init}$ and $\mathrm{VDLS}_{random\_init}$ algorithms are competitive with the VDLS algorithm in solving the instances of the first and second sets. They also significantly outperform the PLTS algorithm in solving most of the 90 instances, indicating that the proposed local search method exhibits a good robustness, as the method can yield high-quality results from various initial solutions.

\section{Conclusion}
\label{sec_Con}

In this paper, we propose an effective Variable Depth Local Search (VDLS) algorithm for solving the Budgeted Maximum Coverage Problem (BMCP), which is a generalization of the NP-hard set cover problem and 0-1 knapsack problem, and a variant problem of the NP-hard Set-Union Knaspsack Problem (SUKP). VDLS first generates an initial solution by a constructive greedy algorithm, then improves the solution iteratively by a partial depth-first search method, that allows the algorithm to explore the solution space widely and deeply. We also define an effective neighbour structure for the BMCP to improve the performance and the efficiency of the local search process. The branching heuristic based on the neighbour structure allows VDLS to explore the solution space efficiently, by abandoning low-quality branch nodes.

We compare the VDLS with the general CPLEX solver and the best-performing heuristic algorithm called PLTS in solving the 30 public BMCP benchmark instances, as well as the 60 proposed new instances in relatively large scales and with complex structural features. The results show that the VDLS outperforms the CPLEX solver and the PLTS significantly, indicating its excellent performance for solving the BMCP. The improvement of VDLS over PLTS is more obviously on the 60 new instances, indicating that our method is more effective for solving large and complex instances, and our designed instances can distinguish the performance of different algorithms better. Moreover, we also do sufficient ablation studies on the branch strategy and the initialization method. The results demonstrate that our proposed neighbour structure is effective and efficient for the VDLS to solve the BMCP, and the partial depth-first search method in the VDLS is robust, that it can yield high-quality solutions with various initial solutions.

The neighbour structure we defined could also be applied to other local search algorithms for the BMCP. Moreover, since our proposed variable depth local search approach is effective in solving the BMCP, in future works we consider applying this method for other combinatorial optimization problems, such as various variants of the knapsack problems and the set cover problems. 

\section*{List of Abbreviations}

\begin{itemize}
    \item \textbf{BMCP} The Budgeted Maximum Coverage Problem.
    \item \textbf{VDLS} The proposed Variable Depth Local Search algorithm.
    \item \textbf{SUKP} The Set-Union Knapsack Problem.
    \item \textbf{PLTS} The baseline heuristic, probability learning based tabu search algorithm.
    \item \textbf{LB} The lower bound calculated by the CPLEX solver.
    \item \textbf{UB} The upper bound calculated by the CPLEX solver.
\end{itemize}

\section*{Declarations}

\begin{itemize}
    \item \textbf{Availability of data and material} 
    
    The datasets generated during the current study are available at https://github.com/JHL-HUST/VDLS/.
    \item \textbf{Competing interests}
    
    The authors have no relevant financial or non-financial interests to disclose.
    \item \textbf{Funding}
    
    No funds, grants, or other support was received.
    \item \textbf{Authors' contributions}
    
    Methodology: Jianrong Zhou; Formal analysis and investigation: Jianrong Zhou, Jiongzhi Zheng, Kun He; Writing - original draft preparation: Jianrong Zhou, Jiongzhi Zheng; Writing - review and editing: Jianrong Zhou, Jiongzhi Zheng, Kun He; Supervision: Kun He.
    
    \item \textbf{Acknowledgements}
    
    None
\end{itemize}



\bibliography{sn-bibliography}


\end{document}